\documentclass[copyright]{eptcs}
\providecommand{\publicationstatus}{}
\providecommand{\event}{ICE 2013}

\usepackage{amssymb}
\usepackage{amsmath}
\usepackage{tikz}
\usepackage{listings}
\usepackage{algorithm}
\usepackage[noend]{algorithmic}

\lstdefinelanguage{ABS}{
 language=Java,
 deletekeywords={void,long,int,byte,boolean,true,false}, 
 morekeywords={type, data, def, cog, export, import, get, let, then, in, await, assert,suspend,module,from,
delta,uses,adds,modifies,removes,original,core,productline,corefeatures,optionalfeatures,after,when,product,hasAttribute,hasMethod},
}

\newenvironment{coreabsdefnblock}[3][]{ \framebox{\mbox{#2}} \quad #3 \\[0pt]}{}

\newcommand{\coreabskw}[1]{\mathbf{#1}}

\usepackage{stmaryrd}

\newcommand{\absnetsplitdrule}[4][]{{\displaystyle\frac{\begin{array}{l}#2\end{array}}{#3}\quad\absnetsplitdrulename{#4}}}

\newcommand{\absnetsplitpremise}[1]{ #1 \\}
\newenvironment{absnetsplitdefnblock}[3][]{ \framebox{\mbox{#2}} \quad #3 \\[0pt]}{}

\newcommand{\absnetsplitnt}[1]{\mathit{#1}}
\newcommand{\absnetsplitmv}[1]{\mathit{#1}}

\newcommand{\absnetsplitsym}[1]{#1}

\newcommand{\absnetsplitdrulename}[1]{\textsc{#1}}

\usepackage{stmaryrd}

\newcommand{\absnetsplitdruleosXXnetstXXstXXstXXredXXnetst}[1]{\absnetsplitdrule[#1]{%
\absnetsplitpremise{\absnetsplitnt{netst}  \longrightarrow  \absnetsplitnt{netst'}}%
}{
\absnetsplitnt{netst} \, \absnetsplitnt{st}  \longrightarrow  \absnetsplitnt{netst'} \, \absnetsplitnt{st}}{%
{\absnetsplitdrulename{os\_netst\_st\_st\_red\_netst}}{}%
}}

\newcommand{\absnetsplitdruleosXXnetstXXstXXstXXredXXst}[1]{\absnetsplitdrule[#1]{%
\absnetsplitpremise{\absnetsplitnt{st}  \longrightarrow  \absnetsplitnt{st'}}%
}{
\absnetsplitnt{netst} \, \absnetsplitnt{st}  \longrightarrow  \absnetsplitnt{netst} \, \absnetsplitnt{st'}}{%
{\absnetsplitdrulename{os\_netst\_st\_st\_red\_st}}{}%
}}

\newcommand{\absnetsplitdruleosXXnetstXXstXXstXXredXXstXXoutXXnetstXXin}[1]{\absnetsplitdrule[#1]{%
\absnetsplitpremise{ \absnetsplitnt{st}  \stackrel{  \overline{ \absnetsplitnt{labeldata} }  }{ \longrightarrow }  \absnetsplitnt{st'} }%
\absnetsplitpremise{ \absnetsplitnt{netst}  \stackrel{  \absnetsplitnt{labeldata}  }{ \longrightarrow }  \absnetsplitnt{netst'} }%
}{
\absnetsplitnt{netst} \, \absnetsplitnt{st}  \longrightarrow  \absnetsplitnt{netst'} \, \absnetsplitnt{st'}}{%
{\absnetsplitdrulename{os\_netst\_st\_st\_red\_st\_out\_netst\_in}}{}%
}}

\newcommand{\absnetsplitdruleosXXnetstXXstXXstXXredXXnetstXXoutXXstXXin}[1]{\absnetsplitdrule[#1]{%
\absnetsplitpremise{ \absnetsplitnt{netst}  \stackrel{  \overline{ \absnetsplitnt{labeldata} }  }{ \longrightarrow }  \absnetsplitnt{netst'} }%
\absnetsplitpremise{ \absnetsplitnt{st}  \stackrel{  \absnetsplitnt{labeldata}  }{ \longrightarrow }  \absnetsplitnt{st'} }%
}{
\absnetsplitnt{netst} \, \absnetsplitnt{st}  \longrightarrow  \absnetsplitnt{netst'} \, \absnetsplitnt{st'}}{%
{\absnetsplitdrulename{os\_netst\_st\_st\_red\_netst\_out\_st\_in}}{}%
}}

\newcommand{\absnetsplitdrulenetXXnetXXtableXXsend}[1]{\absnetsplitdrule[#1]{%
\absnetsplitpremise{\absnetsplitnt{u'}  \neq  \absnetsplitnt{u}}%
\absnetsplitpremise{ \mathit{Q}  \xrightarrow{ \mathrm{enqueue} \, \absnetsplitsym{(}  \textsc{Table} \, \absnetsplitsym{(}  \tau  \absnetsplitsym{)}  \absnetsplitsym{)} }  \mathit{Q}' }%
}{
\mathsf{nd} \, \absnetsplitsym{(}  \absnetsplitnt{u}  \absnetsplitsym{,}  \tau  \absnetsplitsym{)} \, \mathsf{ar} \, \absnetsplitsym{(}  \absnetsplitnt{u}  \absnetsplitsym{,}  \mathit{Q}  \absnetsplitsym{,}  \absnetsplitnt{u'}  \absnetsplitsym{)}  \rightarrow  \mathsf{nd} \, \absnetsplitsym{(}  \absnetsplitnt{u}  \absnetsplitsym{,}  \tau  \absnetsplitsym{)} \, \mathsf{ar} \, \absnetsplitsym{(}  \absnetsplitnt{u}  \absnetsplitsym{,}  \mathit{Q}'  \absnetsplitsym{,}  \absnetsplitnt{u}  \absnetsplitsym{)}}{%
{\absnetsplitdrulename{net\_net\_table\_send}}{}%
}}

\newcommand{\absnetsplitdrulenetXXnetXXtableXXrecv}[1]{\absnetsplitdrule[#1]{%
\absnetsplitpremise{ \mathit{Q}  \xrightarrow{ \mathrm{dequeue} \, \absnetsplitsym{(}  \textsc{Table} \, \absnetsplitsym{(}  \tau'  \absnetsplitsym{)}  \absnetsplitsym{)} }  \mathit{Q}' }%
\absnetsplitpremise{ \tau  \xrightarrow{ \mathrm{update} \, \absnetsplitsym{(}  \tau'  \absnetsplitsym{,}  \absnetsplitnt{u'}  \absnetsplitsym{)} }  \tau'' }%
}{
\mathsf{ar} \, \absnetsplitsym{(}  \absnetsplitnt{u'}  \absnetsplitsym{,}  \mathit{Q}  \absnetsplitsym{,}  \absnetsplitnt{u}  \absnetsplitsym{)} \, \mathsf{nd} \, \absnetsplitsym{(}  \absnetsplitnt{u}  \absnetsplitsym{,}  \tau  \absnetsplitsym{)}  \rightarrow  \mathsf{ar} \, \absnetsplitsym{(}  \absnetsplitnt{u'}  \absnetsplitsym{,}  \mathit{Q}'  \absnetsplitsym{,}  \absnetsplitnt{u}  \absnetsplitsym{)} \, \mathsf{nd} \, \absnetsplitsym{(}  \absnetsplitnt{u}  \absnetsplitsym{,}  \tau''  \absnetsplitsym{)}}{%
{\absnetsplitdrulename{net\_net\_table\_recv}}{}%
}}

\newcommand{\absnetsplitdrulenetXXnetXXrouteXXfurther}[1]{\absnetsplitdrule[#1]{%
\absnetsplitpremise{ \mathit{Q}_{{\mathrm{1}}}  \xrightarrow{ \mathrm{dequeue} \, \absnetsplitsym{(}  \absnetsplitnt{msg}  \absnetsplitsym{)} }  \mathit{Q}'_{{\mathrm{1}}} }%
\absnetsplitpremise{\mathrm{dest} \, \absnetsplitsym{(}  \absnetsplitnt{msg}  \absnetsplitsym{)}  \absnetsplitsym{=}  \absnetsplitnt{o}}%
\absnetsplitpremise{\absnetsplitnt{o} \, \notin \, \tau}%
\absnetsplitpremise{\mathrm{next} \, \absnetsplitsym{(}  \tau  \absnetsplitsym{,}  \absnetsplitnt{o}  \absnetsplitsym{,}  \absnetsplitnt{u}  \absnetsplitsym{)}  \absnetsplitsym{=}  \absnetsplitnt{u''}}%
\absnetsplitpremise{ \mathit{Q}_{{\mathrm{2}}}  \xrightarrow{ \mathrm{enqueue} \, \absnetsplitsym{(}  \absnetsplitnt{msg}  \absnetsplitsym{)} }  \mathit{Q}'_{{\mathrm{2}}} }%
}{
\mathsf{ar} \, \absnetsplitsym{(}  \absnetsplitnt{u'}  \absnetsplitsym{,}  \mathit{Q}_{{\mathrm{1}}}  \absnetsplitsym{,}  \absnetsplitnt{u}  \absnetsplitsym{)} \, \mathsf{nd} \, \absnetsplitsym{(}  \absnetsplitnt{u}  \absnetsplitsym{,}  \tau  \absnetsplitsym{)} \, \mathsf{ar} \, \absnetsplitsym{(}  \absnetsplitnt{u}  \absnetsplitsym{,}  \mathit{Q}_{{\mathrm{2}}}  \absnetsplitsym{,}  \absnetsplitnt{u''}  \absnetsplitsym{)}  \rightarrow  \mathsf{ar} \, \absnetsplitsym{(}  \absnetsplitnt{u'}  \absnetsplitsym{,}  \mathit{Q}'_{{\mathrm{1}}}  \absnetsplitsym{,}  \absnetsplitnt{u}  \absnetsplitsym{)} \, \mathsf{nd} \, \absnetsplitsym{(}  \absnetsplitnt{u}  \absnetsplitsym{,}  \tau  \absnetsplitsym{)} \, \mathsf{ar} \, \absnetsplitsym{(}  \absnetsplitnt{u}  \absnetsplitsym{,}  \mathit{Q}'_{{\mathrm{2}}}  \absnetsplitsym{,}  \absnetsplitnt{u''}  \absnetsplitsym{)}}{%
{\absnetsplitdrulename{net\_net\_route\_further}}{}%
}}

\newcommand{\absnetsplitdrulenetXXnetXXlXXobjectXXsendXXin}[1]{\absnetsplitdrule[#1]{%
\absnetsplitpremise{\absnetsplitnt{o} \, \in \, \tau}%
\absnetsplitpremise{\absnetsplitnt{u'}  \neq  \absnetsplitnt{u}}%
\absnetsplitpremise{ \tau  \xrightarrow{ \mathrm{replace} \, \absnetsplitsym{(}  \absnetsplitnt{o}  \absnetsplitsym{,}  \absnetsplitnt{u'}  \absnetsplitsym{,}  \absnetsplitsym{1}  \absnetsplitsym{)} }  \tau' }%
\absnetsplitpremise{ \mathit{Q}  \xrightarrow{ \mathrm{enqueue} \, \absnetsplitsym{(}  \textsc{Object} \, \absnetsplitsym{(}  \absnetsplitnt{object}  \absnetsplitsym{)}  \absnetsplitsym{)} }  \mathit{Q}' }%
}{
 \mathsf{nd} \, \absnetsplitsym{(}  \absnetsplitnt{u}  \absnetsplitsym{,}  \tau  \absnetsplitsym{)} \, \mathsf{ar} \, \absnetsplitsym{(}  \absnetsplitnt{u}  \absnetsplitsym{,}  \mathit{Q}  \absnetsplitsym{,}  \absnetsplitnt{u'}  \absnetsplitsym{)}  \stackrel{  \mathrm{mv} \, \absnetsplitsym{(}  \absnetsplitnt{object}  \absnetsplitsym{)}  }{ \rightarrow }  \mathsf{nd} \, \absnetsplitsym{(}  \absnetsplitnt{u}  \absnetsplitsym{,}  \tau'  \absnetsplitsym{)} \, \mathsf{ar} \, \absnetsplitsym{(}  \absnetsplitnt{u}  \absnetsplitsym{,}  \mathit{Q}'  \absnetsplitsym{,}  \absnetsplitnt{u'}  \absnetsplitsym{)} }{%
{\absnetsplitdrulename{net\_net\_l\_object\_send\_in}}{}%
}}

\newcommand{\absnetsplitdrulenetXXnetXXlXXmsgXXsendXXin}[1]{\absnetsplitdrule[#1]{%
\absnetsplitpremise{\absnetsplitnt{o} \, \in \, \tau}%
\absnetsplitpremise{\mathrm{dest} \, \absnetsplitsym{(}  \absnetsplitnt{msg}  \absnetsplitsym{)}  \absnetsplitsym{=}  \absnetsplitnt{o'}}%
\absnetsplitpremise{\mathrm{next} \, \absnetsplitsym{(}  \tau  \absnetsplitsym{,}  \absnetsplitnt{o'}  \absnetsplitsym{,}  \absnetsplitnt{u}  \absnetsplitsym{)}  \absnetsplitsym{=}  \absnetsplitnt{u'}}%
\absnetsplitpremise{ \mathit{Q}  \xrightarrow{ \mathrm{enqueue} \, \absnetsplitsym{(}  \absnetsplitnt{msg}  \absnetsplitsym{)} }  \mathit{Q}' }%
}{
 \mathsf{nd} \, \absnetsplitsym{(}  \absnetsplitnt{u}  \absnetsplitsym{,}  \tau  \absnetsplitsym{)} \, \mathsf{ar} \, \absnetsplitsym{(}  \absnetsplitnt{u}  \absnetsplitsym{,}  \mathit{Q}  \absnetsplitsym{,}  \absnetsplitnt{u'}  \absnetsplitsym{)}  \stackrel{  \mathrm{tr} \, \absnetsplitsym{(}  \absnetsplitnt{o}  \absnetsplitsym{,}  \absnetsplitnt{msg}  \absnetsplitsym{)}  }{ \rightarrow }  \mathsf{nd} \, \absnetsplitsym{(}  \absnetsplitnt{u}  \absnetsplitsym{,}  \tau  \absnetsplitsym{)} \, \mathsf{ar} \, \absnetsplitsym{(}  \absnetsplitnt{u}  \absnetsplitsym{,}  \mathit{Q}'  \absnetsplitsym{,}  \absnetsplitnt{u'}  \absnetsplitsym{)} }{%
{\absnetsplitdrulename{net\_net\_l\_msg\_send\_in}}{}%
}}

\newcommand{\absnetsplitdrulenetXXnetXXlXXobjectXXrecvXXout}[1]{\absnetsplitdrule[#1]{%
\absnetsplitpremise{\mathrm{id} \, \absnetsplitsym{(}  \absnetsplitnt{object}  \absnetsplitsym{)}  \absnetsplitsym{=}  \absnetsplitnt{o}}%
\absnetsplitpremise{ \mathit{Q}  \xrightarrow{ \mathrm{dequeue} \, \absnetsplitsym{(}  \textsc{Object} \, \absnetsplitsym{(}  \absnetsplitnt{object}  \absnetsplitsym{)}  \absnetsplitsym{)} }  \mathit{Q}' }%
\absnetsplitpremise{ \tau  \xrightarrow{ \mathrm{replace} \, \absnetsplitsym{(}  \absnetsplitnt{o}  \absnetsplitsym{,}  \absnetsplitnt{u}  \absnetsplitsym{,}  0  \absnetsplitsym{)} }  \tau' }%
}{
 \mathsf{ar} \, \absnetsplitsym{(}  \absnetsplitnt{u'}  \absnetsplitsym{,}  \mathit{Q}  \absnetsplitsym{,}  \absnetsplitnt{u}  \absnetsplitsym{)} \, \mathsf{nd} \, \absnetsplitsym{(}  \absnetsplitnt{u}  \absnetsplitsym{,}  \tau  \absnetsplitsym{)}  \stackrel{  \overline{ \mathrm{mv} \, \absnetsplitsym{(}  \absnetsplitnt{object}  \absnetsplitsym{)} }  }{ \rightarrow }  \mathsf{ar} \, \absnetsplitsym{(}  \absnetsplitnt{u'}  \absnetsplitsym{,}  \mathit{Q}'  \absnetsplitsym{,}  \absnetsplitnt{u}  \absnetsplitsym{)} \, \mathsf{nd} \, \absnetsplitsym{(}  \absnetsplitnt{u}  \absnetsplitsym{,}  \tau'  \absnetsplitsym{)} }{%
{\absnetsplitdrulename{net\_net\_l\_object\_recv\_out}}{}%
}}

\newcommand{\absnetsplitdrulenetXXnetXXlXXmsgXXrecvXXout}[1]{\absnetsplitdrule[#1]{%
\absnetsplitpremise{ \mathit{Q}  \xrightarrow{ \mathrm{dequeue} \, \absnetsplitsym{(}  \absnetsplitnt{msg}  \absnetsplitsym{)} }  \mathit{Q}' }%
\absnetsplitpremise{\mathrm{dest} \, \absnetsplitsym{(}  \absnetsplitnt{msg}  \absnetsplitsym{)}  \absnetsplitsym{=}  \absnetsplitnt{o}}%
\absnetsplitpremise{\absnetsplitnt{o} \, \in \, \tau}%
}{
 \mathsf{ar} \, \absnetsplitsym{(}  \absnetsplitnt{u'}  \absnetsplitsym{,}  \mathit{Q}  \absnetsplitsym{,}  \absnetsplitnt{u}  \absnetsplitsym{)} \, \mathsf{nd} \, \absnetsplitsym{(}  \absnetsplitnt{u}  \absnetsplitsym{,}  \tau  \absnetsplitsym{)}  \stackrel{  \overline{ \mathrm{tr} \, \absnetsplitsym{(}  \absnetsplitnt{o}  \absnetsplitsym{,}  \absnetsplitnt{msg}  \absnetsplitsym{)} }  }{ \rightarrow }  \mathsf{ar} \, \absnetsplitsym{(}  \absnetsplitnt{u'}  \absnetsplitsym{,}  \mathit{Q}'  \absnetsplitsym{,}  \absnetsplitnt{u}  \absnetsplitsym{)} \, \mathsf{nd} \, \absnetsplitsym{(}  \absnetsplitnt{u}  \absnetsplitsym{,}  \tau  \absnetsplitsym{)} }{%
{\absnetsplitdrulename{net\_net\_l\_msg\_recv\_out}}{}%
}}

\newcommand{\absnetsplitdrulenetXXnetXXlXXnewXXobjectXXin}[1]{\absnetsplitdrule[#1]{%
\absnetsplitpremise{\mathrm{fresh} \, \absnetsplitsym{(}  \absnetsplitnt{o'}  \absnetsplitsym{)}}%
\absnetsplitpremise{\absnetsplitnt{o} \, \in \, \tau}%
\absnetsplitpremise{ \tau  \xrightarrow{ \mathrm{register} \, \absnetsplitsym{(}  \absnetsplitnt{o'}  \absnetsplitsym{,}  \absnetsplitnt{u}  \absnetsplitsym{,}  0  \absnetsplitsym{)} }  \tau' }%
}{
 \mathsf{nd} \, \absnetsplitsym{(}  \absnetsplitnt{u}  \absnetsplitsym{,}  \tau  \absnetsplitsym{)}  \stackrel{  \mathrm{rg} \, \absnetsplitsym{(}  \absnetsplitnt{o}  \absnetsplitsym{,}  \absnetsplitnt{o'}  \absnetsplitsym{)}  }{ \rightarrow }  \mathsf{nd} \, \absnetsplitsym{(}  \absnetsplitnt{u}  \absnetsplitsym{,}  \tau'  \absnetsplitsym{)} }{%
{\absnetsplitdrulename{net\_net\_l\_new\_object\_in}}{}%
}}

\usetikzlibrary{shapes,decorations,matrix,arrows,automata,plotmarks}
\tikzstyle{nd} = [draw=black, fill=white, very thick,rectangle, rounded corners, inner xsep=50pt, inner ysep=65pt]
\tikzstyle{nc} = [draw=black, fill=white, very thick,rectangle, rounded corners, inner xsep=40pt, inner ysep=22pt]
\tikzstyle{obj} = [draw=gray, fill=gray!20, very thick,rectangle, rounded corners, inner xsep=5pt, inner ysep=6pt]
\tikzstyle{ol} = [draw=black, fill=white, very thick,rectangle, rounded corners, inner xsep=40pt, inner ysep=22pt]
\tikzstyle{rt} = [draw=black, fill=white, very thick,rectangle, rounded corners, inner xsep=35pt, inner ysep=10pt]
\tikzstyle{ndtitle} =[text=black]

\title{ABS-NET: Fully Decentralized Runtime Adaptation for Distributed Objects}

\author{Karl Palmskog \qquad Mads Dam \qquad Andreas Lundblad
\institute{School of Computer Science and Communication\\
KTH Royal Institute of Technology\\
Stockholm, Sweden}
\email{\{palmskog,mfd,landreas\}@kth.se}
\and
Ali Jafari
\institute{School of Computer Science\\
Reykjavik University\\
Reykjavik, Iceland}
\email{ali11@ru.is}
}
\def\titlerunning{ABS-NET: Fully Decentralized Runtime Adaptation for Distributed Objects}
\def\authorrunning{K. Palmskog, M. Dam, A. Lundblad \& A. Jafari}

\begin{document}
\maketitle

\begin{abstract}
We present a formalized, fully decentralized runtime semantics for a core subset of ABS, a language and framework for modelling distributed object-oriented systems. The semantics incorporates an abstract graph representation of a network infrastructure, with network endpoints represented as graph nodes, and links as arcs with buffers, corresponding to OSI layer 2 interconnects. The key problem we wish to address is how to allocate computational tasks to nodes so that certain performance objectives are met. To this end, we use the semantics as a foundation for performing network-adaptive task execution via object migration between nodes. Adaptability is analyzed in terms of three Quality of Service objectives: node load, arc load and message latency. We have implemented the key parts of our semantics in a simulator and evaluated how well objectives are achieved for some application-relevant choices of network topology, migration procedure and ABS program. The evaluation suggests that it is feasible in a decentralized setting to continually meet both the objective of a node-balanced task allocation and make headway towards minimizing communication, and thus arc load and message latency.
\end{abstract}

\section{Introduction}
An important problem, made more relevant by recent interest in cloud computing, is how to decouple computational processes from the underlying physical infrastructure on which they execute. One motivation for such decoupling is to free applications from handling resource allocation issues, which can instead be taken care of in a provably correct and transparent fashion using generic, application-independent mechanisms \cite{Dam13}. Potentially, tasks can then be performed at the physical machine most suited at the moment, continually meeting global system requirements such as utilization and power consumption, or task-local requirements such as a response time.

We consider the problem of runtime adaptation of tasks in the context of a core subset of ABS \cite{johnsen10fmco}, a language for modelling distributed object-oriented systems developed in the EU FP7 HATS project. In ongoing work \cite{Dam13, DamP13} we are developing network-aware semantics for different fragments of ABS with some novel features. Specifically, we let objects execute on network nodes connected point-to-point using asynchronous message passing links, and show how location independent routing in such a setting can be used to support efficient, transparent, and robust (lock-free) object migration. Here, we examine how adaptation can be performed in such a model by a controller process running on each node.

To enable precise reasoning and experiments on adaptability, we define three central Quality of Services (QoS) objectives against which a solution for runtime adaptation in our context can be assessed: node load, arc load and message latency. We abstract from many practical, implementation-level concerns when interpreting these objectives in our setting. The load for a node is the number of active tasks running on it. The load for an arc is the number of messages traversing the arc. The latency for a message is the number of hops needed to reach its destination. We then restrict our consideration of adaptability to the problem of how and when to migrate objects to achieve the objectives as well as possible, given a specific network topology, ABS program, and node-local procedure for managing migrations. Using a simulator which implements the key parts of our semantics, we have investigated how well objectives are met for some application-relevant choices of network topologies, programs and migration procedures.

Section \ref{abs-background} and \ref{net-model} describe the ABS language and our novel ABS-NET semantics for execution of ABS programs in a network. Section \ref{adaptation} describes our approach to runtime adaptation via object migration. Section \ref{simulator} describes the simulator, our benchmark scenarios, and simulation results; Section \ref{conclusions} concludes.

\section{ABS Background}
\label{abs-background}
ABS \cite{FullABS} is a language and framework for modelling distributed object-oriented systems, developed in the EU FP7 HATS project. Core ABS \cite{johnsen10fmco} is a language which contains the main features of ABS: a functional level for expressing data structures and side-effect free internal computations of objects, and an object level for expressing concurrent objects, and communication among such objects via method invocation. The object level defines syntax---reminiscent of Java's---for interfaces, classes, methods, object creation and method calls. The object level is accompanied by a type system and an operational semantics which preserves well-typing. One consequence of well-typing is that many runtime errors are ruled out for type-checked programs; when an object makes a call to a method $m$ using an object identifier $o$, there always exists an object associated with $o$, which is an instance of a class which implements $m$. ABS uses placeholders in the form of futures for the result of method calls, allowing a caller to avoid blocking until the result value is actually required. In the variant of Core ABS we consider, a single object is the unit of concurrency, as in the variant of Albert et al. \cite{Schlatte11}. This means that objects at runtime can be viewed as actors, communicating between themselves only via asynchronous message passing.

An example of an ABS interface with implementing classes is given below. The \texttt{CastNode} interface defines a method \texttt{aggregate}, which, when called on an object, performs a convergecast operation in the object-reference binary tree rooted at that object. Specifically, this means that if an object implementing \texttt{CastNode} is a leaf in the tree (an instance of \texttt{LeafNode}), it simply returns a locally known integer, but if the object has child nodes in the tree (an instance of \texttt{BranchNode}), \texttt{aggregate} is called on both of those objects and the results are added to the local integer and returned. In this way, the \texttt{aggregate} method for the object $o$ always returns the aggregate (sum) of all local values in the binary tree of objects rooted at $o$. The variables \texttt{fLeft} and \texttt{fRight} in the implementation of \texttt{aggregate} hold the placeholders (futures) for integers that result from the asynchronous method calls. The values are then retrieved through the .$\coreabskw{get}$ operator, which can cause blocking until the method call has finished.

\noindent\begin{minipage}{.40\textwidth}
\begin{lstlisting}
  interface CastNode { 
      Int aggregate(); 
  }
\end{lstlisting}
\end{minipage}\hfill
\begin{minipage}{.75\textwidth}
\begin{lstlisting}
class LeafCastNode(Int val) implements CastNode {
    Int aggregate() { return val; }
}
\end{lstlisting}
\end{minipage}
\begin{lstlisting}           
  class BranchCastNode(Int val, CastNode left, CastNode right) implements CastNode {
     Int aggregate() {
        Fut<Int> fLeft = left!aggregate();
        Fut<Int> fRight = right!aggregate();
        Int aggregateLeft = fLeft.get;
        Int aggregateRight = fRight.get;    
        return val + aggregateLeft + aggregateRight;
      }
  }
\end{lstlisting}

\section{Network Model and Semantics}
\label{net-model}
\begin{figure}[!ht]
  \begin{center}
  \begin{tikzpicture}

    \begin{scope}
      \node [nd] (nd0) { };
      \node[ndtitle, below=0.5pt] at (nd0.north) {node $u_0$};
      \node [nc] [yshift=1cm] (nc0) { };
      \node[ndtitle, below=2pt] at (nc0.north) {node controller};
      \node [ol] [below of=nc0, node distance=2.2cm] (ol0) { };
      \node [rt, above of=ol0, node distance=2cm] (rt0) { };
      \node[ndtitle, below=2pt] at (rt0.north) {routing table};
      \node [obj] [yshift=-1.1cm] (obj1) { $o_1$ };
      \node [obj] [left of=obj1, node distance=0.85cm] (obj0) { $o_0$ };
      \node[ndtitle, below=28pt] at (ol0.north) {interpreter};
    \end{scope}
    
    \begin{scope}[xshift=6.5cm]
      \node [nd] (nd1) { };
      \node[ndtitle, below=0.5pt] at (nd1.north) {node $u_1$};
      \node [nc] [yshift=1cm] (nc1) { };
      \node[ndtitle, below=2pt] at (nc1.north) {node controller};
      \node [ol] [below of=nc1, node distance=2.2cm] (ol1) { };
      \node [rt, above of=ol1, node distance=2cm] (rt1) { };
      \node[ndtitle, below=2pt] at (rt1.north) {routing table};
      \node [obj] [yshift=-1.1cm] (obj2) { $o_2$ };
      \node[ndtitle, below=28pt] at (ol1.north) {interpreter};
    \end{scope}

    \path[<->] (nc0.south) edge[thick] (ol0.north);
    \path[<->] (nc1.south) edge[thick] (ol1.north);
    \path[->] (obj1.east) edge[thick, dashed] (obj2.west);

    \draw[->, very thick] ([yshift=0.5cm]nd0.mid east)
    node (nd0ne) {} 
    node[rectangle, draw=black, fill=white, right of=nd0ne, node distance=1.1cm, inner sep=5pt] (q0) {}  
    node[rectangle, draw=black, fill=white, right of=q0, node distance=0.38cm, inner sep=5pt] (q1) {}
    node[rectangle, draw=black, fill=white, right of=q1, node distance=0.38cm, inner sep=5pt] (q2) {}
    |- ([yshift=0.5cm]nd1.mid west);
    
    \draw[->, very thick] ([yshift=-0.5cm]nd1.mid west) 
    node (nd1nw) {} 
    node[rectangle, draw=black, fill=white, left of=nd1nw, node distance=1.1cm, inner sep=5pt] (q3) {}  
    node[rectangle, draw=black, fill=white, left of=q3, node distance=0.38cm, inner sep=5pt] (q4) {}
    node[rectangle, draw=black, fill=white, left of=q4, node distance=0.38cm, inner sep=5pt] (q5) {}
    |- ([yshift=-0.5cm]nd0.mid east);
    
  \end{tikzpicture}
  \end{center}
  \caption{Nodes, node controllers, and interpreter layers.}
  \label{fig:nodes}
\end{figure}
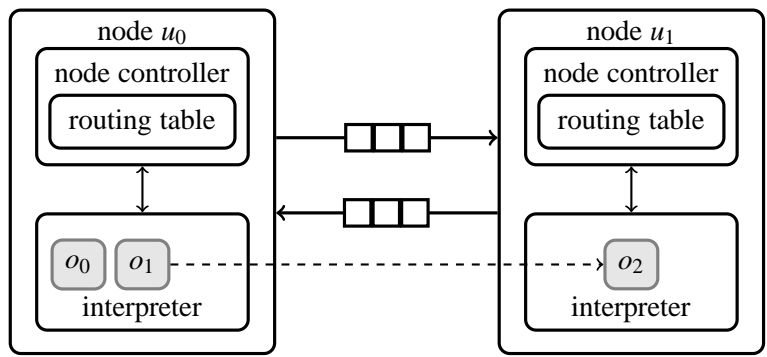

To reason about object adaptability with respect to environmental conditions, we bring selected parts of the infrastructure of a distributed system into our model, namely, network endpoints and links. Endpoints and links are modelled as graph nodes and arcs with FIFO-ordered message queues, respectively. Conceptually, a node consists of an interpreter layer, where local objects reside, and a node controller, which acts as a mediator between the environment and node-local objects, as illustrated in Figure~\ref{fig:nodes}. The dashed arrow in the figure signifies that the identifier of object $o_2$ is known by object $o_1$, allowing $o_1$ to send method invocations to $o_2$. The structure is similar to that used in other programming-language oriented distributed system models, e.g., a proposed semantics for future Erlang \cite{Fredlund10}. Here, the node controller also contains logic for decision-making on adaptability. Seen abstractly, adaptability in this context becomes the problem of deciding when and where to migrate objects to achieve the QoS objectives---with the added constraint that all reallocations must be decided locally at each node.

To achieve location transparency, the basic problem is to route messages correctly between objects that have no prior, mutual knowledge of where they are located. Many solutions have been examined in the literature, including centralized or decentralized location servers, pointer chaining, and broadcast or multicast search. Sewell et al. \cite{Sewell10} discuss many of these solutions, and their relative merits. 

We are developing a novel approach to location transparency based on location-independent (also called name-independent) routing \cite{Dam13}, where the idea is to defer the maintenance of message routes to an explicit routing process executing independently of application-level messaging. Adapted to the approach suggested here, a node controller executing on each network node is responsible for maintaining routing information by exchanging routing tables with adjacent nodes in the network. This allows object migration to be supported in a transparent fashion with only modest extension to the runtime state.

We have defined a new operational semantics of Core ABS programs, in the same rewriting logic style \cite{Clavel02} as the standard semantics, that characterizes task execution of objects located on, and moving between, network nodes. Adaptability features such as routing table exchange and object migration are modelled as nondeterministic events, with the node controller consisting of nothing more than a globally unique identifier and a routing table. We refer to the combination of the Core ABS functional layer, Core ABS object syntax, and our novel operational semantics as ABS-NET. We intend for the semantics to both guide implementation, by defining a baseline for retaining program runtime behaviour similar to Core ABS in a networked, decentralized setting, and provide opportunities for further theoretical analysis of specific adaptability strategies by refinement.

A complete description of the syntax and formal semantics of both Core ABS and ABS-NET is available in an appendix \cite{Palmskog13}, with semantic equivalence explored elsewhere \cite{DamP13}. Below, we give an overview of the ABS-NET network model and semantics, with details on node controller behaviour, which is to an extent agnostic towards the underlying actor environment.

\subsection{Runtime Configurations}
The node controller's relationship with the interpreter layer residing on the node is symbiotic. On one hand, the node controller provides message delivery services and callback functions to obtain new globally unique object identifiers for objects residing in the interpreter layer. On the other hand, the node controller triggers object movement by using callback functions that the interpreter layer makes available. We assume a node controller is aware of the asynchronous links through which it can communicate via message passing with other node controllers. In essence, the aim is that node controllers should be realizable on top of a network with only OSI layer 2 interconnects, meaning that the required primitives for computation and communication can be implemented directly in hardware with high performance. If this is the case, running controllers on top of overlays using higher-layer interconnects such as sockets is also feasible, which is what we do in our simulator.

The global state in ABS-NET is formally a pair $\{ \absnetsplitnt{net} \} \, \{ \absnetsplitnt{cn} \}$. The network part $\{ \absnetsplitnt{net} \}$ is a set of nodes and arcs. In a node $\mathsf{nd} \, \absnetsplitsym{(}  \absnetsplitnt{u}  \absnetsplitsym{,}  \tau  \absnetsplitsym{)}$, $\absnetsplitnt{u}$ is a node identifier (assumed globally unique) and $\tau$ is a routing table, used to route object-related messages in the proper direction. In an arc $\mathsf{ar} \, \absnetsplitsym{(}  \absnetsplitnt{u}  \absnetsplitsym{,}  \mathit{Q}  \absnetsplitsym{,}  \absnetsplitnt{u'}  \absnetsplitsym{)}$, representing a unidirectional link from $\absnetsplitnt{u}$ to $\absnetsplitnt{u'}$, $\mathit{Q}$ is a FIFO-ordered queue of messages. The other part of the global state, $\{ \absnetsplitnt{cn} \}$, is a set of objects, with each object implicitly attached to a node in the network, and able to send and receive messages with the assistance of its host. A message $\absnetsplitnt{msg}$ can be (1) a table message $\textsc{Table} \, \absnetsplitsym{(}  \tau  \absnetsplitsym{)}$, used to pass a routing table $\tau$ from one node to another to update local routes, (2) an object message $\textsc{Object} \, \absnetsplitsym{(}  \absnetsplitnt{object}  \absnetsplitsym{)}$ containing a complete runtime object $\absnetsplitnt{object}$ to facilitate mobility, or (3) an application-level message transmitted from one object to another, which for ABS is either a method invocation (\textsc{Call} message) or the resolved value of a future (\textsc{Future} message). Given an application-level message, the function $\mathtt{dest}(\absnetsplitnt{msg})$ returns the identifier of the intended recipient object, while the function $\mathtt{id}$ returns the identifier of a given runtime object.

The nature of a FIFO queue $\mathit{Q}$ of messages is specified through three functions: $\mathtt{enqueue}$, $\mathtt{dequeue}$ and $\mathtt{first}$. $\mathtt{enqueue}(\mathit{Q}, \absnetsplitnt{msg})$ returns the queue that results when the message $\absnetsplitnt{msg}$ is added to the back of $\mathit{Q}$. If $\mathit{Q}$ is non-empty, $\mathtt{first}(\mathit{Q})$ returns the message at the front of $\mathit{Q}$, and $\mathtt{dequeue}(\mathit{Q})$ returns the queue that results when the front message is removed. For brevity, $\mathtt{enqueue}(\mathit{Q}, \absnetsplitnt{msg}) = \mathit{Q}'$ is defined as a relation {\footnotesize $\mathit{Q}  \xrightarrow{ \mathtt{enqueue} \, \absnetsplitsym{(}  \absnetsplitnt{msg}  \absnetsplitsym{)} }  \mathit{Q}'$}, while the conjunction that $\mathtt{first}(\mathit{Q}) = \absnetsplitnt{msg}$ and $\mathtt{dequeue}(\mathit{Q}) = \mathit{Q}'$ is defined as {\footnotesize $\mathit{Q}  \xrightarrow{ \mathtt{dequeue} \, \absnetsplitsym{(}  \absnetsplitnt{msg} \absnetsplitsym{)} }  \mathit{Q}'$}.

The nature of a routing table is specified through the functions $\mathtt{update}$, $\mathtt{next}$, $\mathtt{register}$ and $\mathtt{replace}$, and an infix operator $\in$. The function $\mathtt{update}$ takes three arguments: the routing table $\tau$ of the current node, the node identifier $\absnetsplitmv{u'}$ of the adjacent node, and the routing table $\tau'$ of the adjacent node. The function returns a routing table $\tau''$, which incorporates the routes from $\tau'$ into $\tau$ if appropriate, with the constraint that all such routes must go through the node $\absnetsplitmv{u'}$. For brevity, $\mathtt{update}(\tau, \absnetsplitmv{u'}, \tau') = \tau''$ is defined as a relation {\footnotesize $\tau  \xrightarrow{ \mathtt{update} \, \absnetsplitsym{(}  \tau'  \absnetsplitsym{,}  \absnetsplitnt{u'}  \absnetsplitsym{)} }  \tau''$}. The function $\mathtt{next}$ takes three arguments: the routing table $\tau$ of the current node, the object identifier $\absnetsplitmv{o'}$ of the node we want the next hop for, and the default hop $\absnetsplitmv{u}$, which is the identifier of the current node. The function returns the node identifier $\absnetsplitmv{u'}$ which is the next hop of $\absnetsplitmv{o'}$ according to the table. The function $\mathtt{register}$ takes four arguments: the routing table $\tau$ of the current node, the object identifier $\absnetsplitmv{o'}$ of the object we want to add a route for, the node identifier $\absnetsplitmv{u}$ of a neighbour node (usually self) which is the next hop, and a non-negative integer $k$ for the distance to the object (in all instances in the rules, it is 0). The function returns a routing table $\tau'$ which incorporates the new route. For brevity, $\mathtt{register}(\tau, \absnetsplitmv{o'}, \absnetsplitmv{u}, k) = \tau'$ is defined as a relation {\footnotesize $\tau  \xrightarrow{ \mathtt{register} \, \absnetsplitsym{(}  \absnetsplitnt{o'}  \absnetsplitsym{,}  \absnetsplitnt{u}  \absnetsplitsym{,}  k  \absnetsplitsym{)} }  \tau'$}. The function $\mathtt{replace}$ takes four arguments (of the same type as $\mathtt{register}$): the routing table $\tau$ of the current node, the object identifier $\absnetsplitmv{o'}$ of the object we want to replace the route for, the node identifier $\absnetsplitmv{u}$ of a neighbour node which is the next hop, and a natural number $k$ for the distance to the object. The function returns a routing table $\tau'$ which has removed any existing routes for $\absnetsplitmv{o'}$ and added the route given. For brevity, $\mathtt{replace}(\tau, \absnetsplitmv{o'}, \absnetsplitmv{u}, k) = \tau'$ is defined as a relation {\footnotesize $\tau  \xrightarrow{ \mathtt{replace} \, \absnetsplitsym{(}  \absnetsplitnt{o}  \absnetsplitsym{,}  \absnetsplitnt{u}  \absnetsplitsym{,}  k  \absnetsplitsym{)} }  \tau'$}. The claim $\absnetsplitmv{o} \in \tau$, with a node identifier $\absnetsplitmv{u}$ given by the context, means that, according to $\tau$, the object with identifier $\absnetsplitmv{o}$ is located on the node $\absnetsplitmv{u}$.

The two parts of the global state can evolve jointly by performing synchronized labelled transitions, but also separately without exchanging information. The rules for such synchronization and separate evolution are shown in Figure~\ref{fig:node-object-rules}. A label $\alpha$ is either $\mathtt{mv}(\absnetsplitnt{object})$ (moving an object), $\mathtt{rg}(\absnetsplitmv{o}, \absnetsplitmv{o'})$ (registering a new object identifier), or $\mathtt{tr}(\absnetsplitmv{o}, \absnetsplitnt{msg})$ (transporting a message). Intuitively, a label with an overline means that information is outgoing or being sent, while a label without overline means information is incoming or being received. An ABS-NET execution of an ABS program is a possibly infinite sequence of global states, such that the transition between a previous state and the next is valid. The program is not explicitly represented in a state, since it assumed to always be available unaltered at all nodes. The network topology is static during an execution, and we do not consider failures such as message losses.

\renewcommand{\absnetsplitdruleosXXnetstXXstXXstXXredXXnetst}[1]{\absnetsplitdrule[#1]{%
\absnetsplitpremise{{\absnetsplitdrulename{(Net-Red)}}}%
\absnetsplitpremise{\{ \absnetsplitnt{net} \}  \rightarrow  \{ \absnetsplitnt{net'} \}}%
}{
\{ \absnetsplitnt{net} \} \,  \{  \absnetsplitnt{cn}  \}  \rightarrow  \{ \absnetsplitnt{net'} \} \,  \{  \absnetsplitnt{cn}  \}}{%
{}%
}}

\renewcommand{\absnetsplitdruleosXXnetstXXstXXstXXredXXst}[1]{\absnetsplitdrule[#1]{%
\absnetsplitpremise{{\absnetsplitdrulename{(Cn-Red)}}}%
\absnetsplitpremise{\{  \absnetsplitnt{cn}  \}  \rightarrow \{  \absnetsplitnt{cn'}  \}}%
}{
\{ \absnetsplitnt{net} \} \, \{  \absnetsplitnt{cn}  \}  \rightarrow  \{ \absnetsplitnt{net} \} \, \{  \absnetsplitnt{cn'} \}}{%
{}%
}}

\renewcommand{\absnetsplitdruleosXXnetstXXstXXstXXredXXstXXoutXXnetstXXin}[1]{\absnetsplitdrule[#1]{%
\absnetsplitpremise{{\absnetsplitdrulename{(Cn-Out-Net-In-Red)}}}%
\absnetsplitpremise{ \{  \absnetsplitnt{cn}  \}  \stackrel{  \overline{ \alpha }  }{ \rightarrow }  \{  \absnetsplitnt{cn'}  \} }%
\absnetsplitpremise{ \{ \absnetsplitnt{net} \} \stackrel{  \alpha  }{ \rightarrow }  \{ \absnetsplitnt{net'} \} }%
}{
\{ \absnetsplitnt{net} \} \, \{  \absnetsplitnt{cn}  \} \rightarrow  \{ \absnetsplitnt{net'} \} \, \{  \absnetsplitnt{cn'}  \}}{%
{}%
}}

\renewcommand{\absnetsplitdruleosXXnetstXXstXXstXXredXXnetstXXoutXXstXXin}[1]{\absnetsplitdrule[#1]{%
\absnetsplitpremise{{\absnetsplitdrulename{(Net-Out-Cn-In-Red)}}}%
\absnetsplitpremise{ \{ \absnetsplitnt{net} \}  \stackrel{  \overline{ \alpha }  }{ \rightarrow }  \{ \absnetsplitnt{net'} \} }%
\absnetsplitpremise{ \{  \absnetsplitnt{cn}  \}  \stackrel{  \alpha  }{ \rightarrow }  \{  \absnetsplitnt{cn'}  \} }%
}{
\{ \absnetsplitnt{net} \} \, \{  \absnetsplitnt{cn}  \}  \rightarrow  \{ \absnetsplitnt{net'} \} \, \{  \absnetsplitnt{cn'}  \}}{%
{}%
}}

\begin{figure}[!ht]
\centering{
  {\footnotesize
    $\absnetsplitdruleosXXnetstXXstXXstXXredXXnetst{}$ $\absnetsplitdruleosXXnetstXXstXXstXXredXXst{}$ $\absnetsplitdruleosXXnetstXXstXXstXXredXXstXXoutXXnetstXXin{}$ $\absnetsplitdruleosXXnetstXXstXXstXXredXXnetstXXoutXXstXXin{}$
  }
}
\caption{ABS-NET reduction rules connecting objects and networks.}
\label{fig:node-object-rules}
\end{figure}

Intuitively, transitions by the rule {\footnotesize $\absnetsplitdrulename{Net-Red}$} are driven by node-related events, e.g., timeouts triggering routing table exchanges, or application-level messages being received and routed further. Transitions by {\footnotesize $\absnetsplitdrulename{Cn-Red}$} are triggered when objects execute ABS program statements that only change internal state. {\footnotesize $\absnetsplitdrulename{Cn-Out-Net-In-Red}$} is used when program execution requires interaction with the environment to proceed (e.g., method calls) and for object migration. {\footnotesize $\absnetsplitdrulename{Net-Out-Cn-In-Red}$} is used when a node transmits objects or application-level messages received through links to the local interpreter layer.

\subsection{Node Controller Behaviour}
\label{controller-rules}

The reduction relation for networks, alluded to Figure~\ref{fig:node-object-rules}, is defined by the rules in Figure~\ref{fig:node-rules}. The rules apply to subsets of nodes and arcs, such that the elements can be rearranged to match the left-hand side. The labelled-transition rules {\footnotesize $\absnetsplitdrulename{Net-Msg-Recv-Out}$}, {\footnotesize $\absnetsplitdrulename{Net-Msg-Send-In}$}, {\footnotesize $\absnetsplitdrulename{Net-Object-Send-In}$} and {\footnotesize $\absnetsplitdrulename{Net-New-Object-In}$}, where a node exchanges information with an object, all use the premise $\absnetsplitmv{o} \in \tau$ to restrict actions to pertain to node-local objects. This is how object location is reflected in ABS-NET. $\mathtt{fresh}(o)$ means that the identifier $o$ is globally unique.

\renewcommand{\absnetsplitdrulenetXXnetXXtableXXsend}[1]{\absnetsplitdrule[#1]{%
\absnetsplitpremise{{\absnetsplitdrulename{(Net-Table-Send)}}}
\absnetsplitpremise{\absnetsplitnt{u'}  \neq  \absnetsplitnt{u}}%
\absnetsplitpremise{\mathit{Q}  \xrightarrow{ \mathtt{enqueue} \, \absnetsplitsym{(}  \textsc{Table} \, \absnetsplitsym{(}  \tau  \absnetsplitsym{)}  \absnetsplitsym{)} }  \mathit{Q}' }%
}{
    \begin{array}{c}
\mathsf{nd} \, \absnetsplitsym{(}  \absnetsplitnt{u}  \absnetsplitsym{,}  \tau  \absnetsplitsym{)} \, \mathsf{ar} \, \absnetsplitsym{(}  \absnetsplitnt{u}  \absnetsplitsym{,}  \mathit{Q}  \absnetsplitsym{,}  \absnetsplitnt{u'}  \absnetsplitsym{)}  \\
\rightarrow  \mathsf{nd} \, \absnetsplitsym{(}  \absnetsplitnt{u}  \absnetsplitsym{,}  \tau  \absnetsplitsym{)} \, \mathsf{ar} \, \absnetsplitsym{(}  \absnetsplitnt{u}  \absnetsplitsym{,}  \mathit{Q}'  \absnetsplitsym{,}  \absnetsplitnt{u'}  \absnetsplitsym{)}
\end{array}
}{%
{}%
}}

\renewcommand{\absnetsplitdrulenetXXnetXXtableXXrecv}[1]{\absnetsplitdrule[#1]{%
\absnetsplitpremise{{\absnetsplitdrulename{(Net-Table-Recv)}}}%
\absnetsplitpremise{ \mathit{Q}  \xrightarrow{ \mathtt{dequeue} \, \absnetsplitsym{(}  \textsc{Table} \, \absnetsplitsym{(}  \tau'  \absnetsplitsym{)}  \absnetsplitsym{)} }  \mathit{Q}' }%
\absnetsplitpremise{ \tau  \xrightarrow{ \mathtt{update} \, \absnetsplitsym{(}  \tau'  \absnetsplitsym{,}  \absnetsplitnt{u'}  \absnetsplitsym{)} }  \tau'' }%
}{
    \begin{array}{c}
\mathsf{ar} \, \absnetsplitsym{(}  \absnetsplitnt{u'}  \absnetsplitsym{,}  \mathit{Q}  \absnetsplitsym{,}  \absnetsplitnt{u}  \absnetsplitsym{)} \, \mathsf{nd} \, \absnetsplitsym{(}  \absnetsplitnt{u}  \absnetsplitsym{,}  \tau  \absnetsplitsym{)} \\ 
\rightarrow  \mathsf{ar} \, \absnetsplitsym{(}  \absnetsplitnt{u'}  \absnetsplitsym{,}  \mathit{Q}'  \absnetsplitsym{,}  \absnetsplitnt{u}  \absnetsplitsym{)} \, \mathsf{nd} \, \absnetsplitsym{(}  \absnetsplitnt{u}  \absnetsplitsym{,}  \tau''  \absnetsplitsym{)}
\end{array}
}{%
{}%
}}

\renewcommand{\absnetsplitdrulenetXXnetXXlXXmsgXXsendXXin}[1]{\absnetsplitdrule[#1]{%
\absnetsplitpremise{{\absnetsplitdrulename{(Net-Msg-Send-In)}}}
\absnetsplitpremise{\absnetsplitnt{o} \, \in \, \tau \quad \mathtt{dest} \, \absnetsplitsym{(}  \absnetsplitnt{msg}  \absnetsplitsym{)}  \absnetsplitsym{=}  \absnetsplitnt{o'}}%
\absnetsplitpremise{\mathtt{next} \, \absnetsplitsym{(}  \tau  \absnetsplitsym{,}  \absnetsplitnt{o'}  \absnetsplitsym{,}  \absnetsplitnt{u}  \absnetsplitsym{)}  \absnetsplitsym{=}  \absnetsplitnt{u'}}%
\absnetsplitpremise{ \mathit{Q}  \xrightarrow{ \mathtt{enqueue} \, \absnetsplitsym{(}  \absnetsplitnt{msg}  \absnetsplitsym{)} }  \mathit{Q}' }%
}{
    \begin{array}{c}
 \mathsf{nd} \, \absnetsplitsym{(}  \absnetsplitnt{u}  \absnetsplitsym{,}  \tau  \absnetsplitsym{)} \, \mathsf{ar} \, \absnetsplitsym{(}  \absnetsplitnt{u}  \absnetsplitsym{,}  \mathit{Q}  \absnetsplitsym{,}  \absnetsplitnt{u'}  \absnetsplitsym{)}\\  
 \stackrel{  \mathtt{tr} \, \absnetsplitsym{(}  \absnetsplitnt{o}  \absnetsplitsym{,}  \absnetsplitnt{msg}  \absnetsplitsym{)}  }{ \rightarrow }  \mathsf{nd} \, \absnetsplitsym{(}  \absnetsplitnt{u}  \absnetsplitsym{,}  \tau  \absnetsplitsym{)} \, \mathsf{ar} \, \absnetsplitsym{(}  \absnetsplitnt{u}  \absnetsplitsym{,}  \mathit{Q}'  \absnetsplitsym{,}  \absnetsplitnt{u'}  \absnetsplitsym{)} 
\end{array}
}{%
{}%
}}

\renewcommand{\absnetsplitdrulenetXXnetXXlXXmsgXXrecvXXout}[1]{\absnetsplitdrule[#1]{%
\absnetsplitpremise{{\absnetsplitdrulename{(Net-Msg-Recv-Out)}}}%[1.1em]
\absnetsplitpremise{ \mathit{Q}  \xrightarrow{ \mathtt{dequeue} \, \absnetsplitsym{(}  \absnetsplitnt{msg}  \absnetsplitsym{)} }  \mathit{Q}' }%
\absnetsplitpremise{\mathtt{dest} \, \absnetsplitsym{(}  \absnetsplitnt{msg}  \absnetsplitsym{)}  \absnetsplitsym{=}  \absnetsplitnt{o} \quad \absnetsplitnt{o} \, \in \, \tau}%
}{
    \begin{array}{c}
 \mathsf{ar} \, \absnetsplitsym{(}  \absnetsplitnt{u'}  \absnetsplitsym{,}  \mathit{Q}  \absnetsplitsym{,}  \absnetsplitnt{u}  \absnetsplitsym{)} \, \mathsf{nd} \, \absnetsplitsym{(}  \absnetsplitnt{u}  \absnetsplitsym{,}  \tau  \absnetsplitsym{)}\\  
 \stackrel{  \overline{ \mathtt{tr} \, \absnetsplitsym{(}  \absnetsplitnt{o}  \absnetsplitsym{,}  \absnetsplitnt{msg}  \absnetsplitsym{)} }  }{ \rightarrow }  \mathsf{ar} \, \absnetsplitsym{(}  \absnetsplitnt{u'}  \absnetsplitsym{,}  \mathit{Q}'  \absnetsplitsym{,}  \absnetsplitnt{u}  \absnetsplitsym{)} \, \mathsf{nd} \, \absnetsplitsym{(}  \absnetsplitnt{u}  \absnetsplitsym{,}  \tau  \absnetsplitsym{)} 
\end{array}
}{%
{}%
}}

\renewcommand{\absnetsplitdrulenetXXnetXXrouteXXfurther}[1]{\absnetsplitdrule[#1]{%
\absnetsplitpremise{{\absnetsplitdrulename{(Net-Route-Further)}}}%
\absnetsplitpremise{ \mathit{Q}_{{\mathrm{1}}}  \xrightarrow{ \mathtt{dequeue} \, \absnetsplitsym{(}  \absnetsplitnt{msg}  \absnetsplitsym{)} }  \mathit{Q}'_{{\mathrm{1}}} \quad \mathtt{dest} \, \absnetsplitsym{(}  \absnetsplitnt{msg}  \absnetsplitsym{)}  \absnetsplitsym{=}  \absnetsplitnt{o} \quad \absnetsplitnt{o} \, \notin \, \tau}%
\absnetsplitpremise{\mathtt{next} \, \absnetsplitsym{(}  \tau  \absnetsplitsym{,}  \absnetsplitnt{o}  \absnetsplitsym{,}  \absnetsplitnt{u}  \absnetsplitsym{)}  \absnetsplitsym{=}  \absnetsplitnt{u''} \quad \mathit{Q}_{{\mathrm{2}}}  \xrightarrow{ \mathtt{enqueue} \, \absnetsplitsym{(}  \absnetsplitnt{msg}  \absnetsplitsym{)} }  \mathit{Q}'_{{\mathrm{2}}} }%
}{
    \begin{array}{c}
\mathsf{ar} \, \absnetsplitsym{(}  \absnetsplitnt{u'}  \absnetsplitsym{,}  \mathit{Q}_{{\mathrm{1}}}  \absnetsplitsym{,}  \absnetsplitnt{u}  \absnetsplitsym{)} \, \mathsf{nd} \, \absnetsplitsym{(}  \absnetsplitnt{u}  \absnetsplitsym{,}  \tau  \absnetsplitsym{)} \, \mathsf{ar} \, \absnetsplitsym{(}  \absnetsplitnt{u}  \absnetsplitsym{,}  \mathit{Q}_{{\mathrm{2}}}  \absnetsplitsym{,}  \absnetsplitnt{u''}  \absnetsplitsym{)} \\ 
\rightarrow  \mathsf{ar} \, \absnetsplitsym{(}  \absnetsplitnt{u'}  \absnetsplitsym{,}  \mathit{Q}'_{{\mathrm{1}}}  \absnetsplitsym{,}  \absnetsplitnt{u}  \absnetsplitsym{)} \, \mathsf{nd} \, \absnetsplitsym{(}  \absnetsplitnt{u}  \absnetsplitsym{,}  \tau  \absnetsplitsym{)} \, \mathsf{ar} \, \absnetsplitsym{(}  \absnetsplitnt{u}  \absnetsplitsym{,}  \mathit{Q}'_{{\mathrm{2}}}  \absnetsplitsym{,}  \absnetsplitnt{u''}  \absnetsplitsym{)}
\end{array}
}{%
{}%
}}

\renewcommand{\absnetsplitdrulenetXXnetXXlXXobjectXXsendXXin}[1]{\absnetsplitdrule[#1]{%
\absnetsplitpremise{{\absnetsplitdrulename{(Net-Object-Send-In)}}}
\absnetsplitpremise{\absnetsplitnt{o} \, \in \, \tau \quad \absnetsplitnt{u'}  \neq  \absnetsplitnt{u}}%
\absnetsplitpremise{ \tau  \xrightarrow{ \mathtt{replace} \, \absnetsplitsym{(}  \absnetsplitnt{o}  \absnetsplitsym{,}  \absnetsplitnt{u'}  \absnetsplitsym{,}  \absnetsplitsym{1}  \absnetsplitsym{)} }  \tau' }%
\absnetsplitpremise{ \mathit{Q}  \xrightarrow{ \mathtt{enqueue} \, \absnetsplitsym{(}  \textsc{Object} \, \absnetsplitsym{(}  \absnetsplitnt{object}  \absnetsplitsym{)}  \absnetsplitsym{)} }  \mathit{Q}' }%
}{
\begin{array}{c}
 \mathsf{nd} \, \absnetsplitsym{(}  \absnetsplitnt{u}  \absnetsplitsym{,}  \tau  \absnetsplitsym{)} \, \mathsf{ar} \, \absnetsplitsym{(}  \absnetsplitnt{u}  \absnetsplitsym{,}  \mathit{Q}  \absnetsplitsym{,}  \absnetsplitnt{u'}  \absnetsplitsym{)} \\ 
\stackrel{  \mathtt{mv} \, \absnetsplitsym{(}  \absnetsplitnt{object}  \absnetsplitsym{)}  }{ \rightarrow }  \mathsf{nd} \, \absnetsplitsym{(}  \absnetsplitnt{u}  \absnetsplitsym{,}  \tau'  \absnetsplitsym{)} \, \mathsf{ar} \, \absnetsplitsym{(}  \absnetsplitnt{u}  \absnetsplitsym{,}  \mathit{Q}'  \absnetsplitsym{,}  \absnetsplitnt{u'}  \absnetsplitsym{)} 
\end{array}
}{%
{}%
}}

\renewcommand{\absnetsplitdrulenetXXnetXXlXXobjectXXrecvXXout}[1]{\absnetsplitdrule[#1]{%
\absnetsplitpremise{{\absnetsplitdrulename{(Net-Object-Recv-Out)}}}%
\absnetsplitpremise{\mathtt{id} \, \absnetsplitsym{(}  \absnetsplitnt{object}  \absnetsplitsym{)}  \absnetsplitsym{=}  \absnetsplitnt{o}}%
\absnetsplitpremise{ \mathit{Q}  \xrightarrow{ \mathtt{dequeue} \, \absnetsplitsym{(}  \textsc{Object} \, \absnetsplitsym{(}  \absnetsplitnt{object}  \absnetsplitsym{)}  \absnetsplitsym{)} }  \mathit{Q}' }%
\absnetsplitpremise{ \tau  \xrightarrow{ \mathtt{replace} \, \absnetsplitsym{(}  \absnetsplitnt{o}  \absnetsplitsym{,}  \absnetsplitnt{u}  \absnetsplitsym{,}  0  \absnetsplitsym{)} }  \tau' }%
}{
    \begin{array}{c}
 \mathsf{ar} \, \absnetsplitsym{(}  \absnetsplitnt{u'}  \absnetsplitsym{,}  \mathit{Q}  \absnetsplitsym{,}  \absnetsplitnt{u}  \absnetsplitsym{)} \, \mathsf{nd} \, \absnetsplitsym{(}  \absnetsplitnt{u}  \absnetsplitsym{,}  \tau  \absnetsplitsym{)}\\  
 \stackrel{  \overline{ \mathtt{mv} \, \absnetsplitsym{(}  \absnetsplitnt{object}  \absnetsplitsym{)} }  }{ \rightarrow }  \mathsf{ar} \, \absnetsplitsym{(}  \absnetsplitnt{u'}  \absnetsplitsym{,}  \mathit{Q}'  \absnetsplitsym{,}  \absnetsplitnt{u}  \absnetsplitsym{)} \, \mathsf{nd} \, \absnetsplitsym{(}  \absnetsplitnt{u}  \absnetsplitsym{,}  \tau'  \absnetsplitsym{)} 
\end{array}
}{%
{}%
}}

\renewcommand{\absnetsplitdrulenetXXnetXXlXXnewXXobjectXXin}[1]{\absnetsplitdrule[#1]{%
\absnetsplitpremise{{\absnetsplitdrulename{(Net-New-Object-In)}}}%
\absnetsplitpremise{\mathtt{fresh} \, \absnetsplitsym{(}  \absnetsplitnt{o'}  \absnetsplitsym{)} \quad \absnetsplitnt{o} \, \in \, \tau}%
\absnetsplitpremise{ \tau  \xrightarrow{ \mathtt{register} \, \absnetsplitsym{(}  \absnetsplitnt{o'}  \absnetsplitsym{,}  \absnetsplitnt{u}  \absnetsplitsym{,}  0  \absnetsplitsym{)} }  \tau' }%
}{
 \mathsf{nd} \, \absnetsplitsym{(}  \absnetsplitnt{u}  \absnetsplitsym{,}  \tau  \absnetsplitsym{)}  \stackrel{  \mathtt{rg} \, \absnetsplitsym{(}  \absnetsplitnt{o}  \absnetsplitsym{,}  \absnetsplitnt{o'}  \absnetsplitsym{)}  }{ \rightarrow }  \mathsf{nd} \, \absnetsplitsym{(}  \absnetsplitnt{u}  \absnetsplitsym{,}  \tau'  \absnetsplitsym{)} }{%
{}%
}}

\begin{figure}[!ht]
\centering{
  {\footnotesize
    $\absnetsplitdrulenetXXnetXXtableXXsend{}$ $\absnetsplitdrulenetXXnetXXtableXXrecv{}$ $\absnetsplitdrulenetXXnetXXlXXmsgXXrecvXXout{}$\\[0.5em]
    $\absnetsplitdrulenetXXnetXXlXXmsgXXsendXXin{}$ $\absnetsplitdrulenetXXnetXXrouteXXfurther{}$\\[0.5em]
    $\absnetsplitdrulenetXXnetXXlXXobjectXXsendXXin{}$ $\absnetsplitdrulenetXXnetXXlXXobjectXXrecvXXout{}$ $\absnetsplitdrulenetXXnetXXlXXnewXXobjectXXin{}$
  } 
}
\caption{Node controller reduction rules.}
\label{fig:node-rules}
\end{figure}

For proper progress in execution, we assume networks are such that (1) there are no dangling arcs referencing non-existent nodes, (2) for every arc between nodes there is an arc in the opposite direction, and (3) every node comes with a self-loop arc, i.e., an arc going from and to the node. Self-loop arcs are important for two reasons. First, it allows us to use the same rules for message passing in both the case where the sender object is at a different node from the receiver object, and where the sender is at the same node as the receiver. Once a message has been put in the self-loop queue, it appears as if it came from some other node, and the rule {\footnotesize $\absnetsplitdrulename{Net-Msg-Recv-Out}$} can be applied. Second, it is not always the case that there is a route to the recipient of a message, because of the possibility of stale routing tables. However, messages must be dealt with somehow, in particular if they are coming from some other node, from which there could be other important messages pending. Hence, they are put in the self-loop queue, i.e., the default next hop of an application-level message is the node itself.

The ABS-NET reduction rules for objects at runtime are deferred to a technical report \cite{Palmskog13}. Compared to Core ABS, the state of an object has been extended with an input and an output queues for asynchronous transfer of application-level messages, and a structure to keep track of resolved future values. In contrast, the standard Core ABS semantics handles resolution and querying of futures in a centralized way. In fact, all rule premises from the standard semantics that pertain to more than object-local state are absent in ABS-NET---its decentralized nature is syntactically apparent.

\section{Adaptation}
\label{adaptation}
We consider three QoS objectives against which runtime adaptation solutions can be assessed: node load, arc load and message latency. In our setting, the definition of node load is simple but coarse grained: the load on a node $u$ is the number of objects located on $u$ with active tasks. One advantage of this measure is that it is an intrinsic property of runtime configurations. We need a model-intrinsic measure of load to enable reasoning at an abstract level about convergence to balanced allocation and that loads stay within a certain range. One disadvantage of the approach is that it fails to take into account the varying use of memory and processing power among tasks. However, in an implementation, a more fine-grained measure of load can be adopted, as long as it is linear in the number of active tasks.

We define the load of a particular arc as the number of messages traversing it per simulated unit of time. Hence, global minimization of arc load means that a minimal number of inter-node messages are sent overall, with respect to the current state of routing tables at nodes. Unless all routing tables are optimal (minimum stretch), however, there is no guarantee that the number of hops, i.e., latency, of a particular object-addressed message is minimal.

In our evaluation of runtime adaptation, we use ABS programs that are nonterminating and cyclical. The motivation is that for adaptations to current conditions to have a chance of conveying benefits, similar conditions must hold in the future. There is no obvious payoff in attempting to adapt when future states are random independently of the current state. 

Although we wish to simultaneously meet all of our QoS objectives fully, we consider node load balancing our primary concern. Load balancing solutions are also relatively well-studied in the literature, making it easier to find a good starting point. Azar et al. \cite{Azar99} consider the problem of achieving balanced allocations in the framework of stochastic processes, where it is viewed as stepwise allocations of balls into bins. They highlight the use of greedy schemes for quickly converging to a ball-to-bin assignment where the maximum number of balls in any bin is minimized. The main drawback of this approach in a distributed setting is the reliance on atomic, single assignments of a ball to a bin at each algorithm step. Even-Dar and Mansour \cite{Mansour05} study load balancing in a distributed setting where allocations are not necessarily done one-at-a-time. They give a distributed algorithm for selfish rerouting that quickly converges to a Nash equilibrium, which corresponds to a balanced resource allocation. However, at each round, locally computing a new allocation requires having exact knowledge of all loads in the system, which is complicated and costly to acquire in the current setting. Berenbrink et al. \cite{Berenbrink07} describe and analyze fully distributed algorithms which require only local knowledge of the total number of resources and the load of one other resource to perform a single task migration step. The algorithms, some of which have attractive expected time for convergence, can be straightforwardly translated to a synchronous, round-based distributed setting and further to a message-passing setting, assuming some inherent synchrony. One important assumption made in the algorithm analysis is that a task can migrate to any other resource in a single concurrent round. For this property to hold, the underlying network graph must be complete, which we do not generally assume. 

A factor in the convergence time is whether neutral moves are allowed, i.e., whether a migration can happen even when, as far as can be told locally, the move does not result in a more balanced allocation but merely an equally good one. For allocations in a sparse network graph where load differences between neighbours are one, there can nevertheless be maximal load differences in the order of the graph diameter, which can be significant. With neutral moves, such allocations can be improved on.

The problem of oscillating behaviour during task balancing can be mitigated by the use of coin flips before finalizing decisions to migrate tasks, as in the algorithms of Berenbrink et al. Oscillation can be worsened by reliance on stale information, but if the information is not \emph{too} stale, oscillation periods can sometimes be bounded \cite{Fischer09}.

The literature on load balancing related to scientific computing contains work on simultaneously optimizing task allocations and communication overhead. For example, Cosenza et al. \cite{Cosenza11} give a distributed load balancing scheme for simulations involving agents moving in space from worker to worker. The scheme, which is validated experimentally, optimizes both worker load and communication overhead between workers, but assumes only a small area of interest for each agent, with agents unable to communicate with other agents outside this area. In the current work, two objects can communicate whenever the identifier of one of them is known to the other, making it harder to minimize communication overhead. Catalyurek et al. \cite{Catalyurek07} describe how to use hypergraph partitioning to minimize both communication volume and migration time of tasks for parallel scientific computations. However, the repartitioning is performed in batch and requires complete knowledge of the data and computations on each node.

At this initial stage of the work, we do not consider the cost of migration itself in terms of messaging and other resources. Hence, we only measure communication in terms of messages exchanged between objects, ignoring overhead in terms of routing and load-related messages.

\section{Simulator}
\label{simulator}
We have evaluated our runtime adaptation approach by developing a simulator for running ABS programs in a network of nodes according to the ABS-NET semantics. We have run the simulator with a variety of network node topologies and object migration procedures on a number of proposed synthetic scenarios defined by ABS programs.

Our simulator's main purposes are to serve as a proof-of-concept for ABS-NET and to allow us to run adaptability case studies with particular programs and topologies. Specifically, we are interested in studying convergence properties of object migration procedures in practice, and in showing that our approach to distributed execution scales to networks with many nodes.

The simulator is implemented in Java. Each node controller is implemented as a Java thread, which communicates with other controllers through TCP sockets, using the KryoNet network library \cite{kryonet}. One reason for the choice of sockets is to enable to scale simulations over several physical machines and a large number of simulated network nodes. All node controllers in the network have a representation of the abstract syntax tree of the ABS program being executed, which is generated from ABS program code by the lexing and parsing frontend shared by most ABS backends. 

As in the conceptual model and the formal semantics, a node controller can have zero or more objects, each having at most one active task. An active task has a reference to the statement currently being executed in the abstract syntax tree. We call an object active if it has an active task. Scheduling of active tasks is done at the node controller level in a round-robin fashion for active objects. More precisely, the scheduler deterministically steps all active tasks, checks for active objects, and then repeats the process on the new set of active tasks.

We implement statement execution by interpretation. The main reason for this choice is to enable easy serialization of objects between executing statements; to get immediate results from load balancing, we must be able to migrate active objects. One drawback of using interpretation is that local execution is slow and resource-demanding compared to execution in the standard ABS backends.

A node controller is associated with a unique TCP port on the host system. Besides a list of neighbour handles, which abstract over underlying sockets, and a list of local objects, the node controller maintains a routing table. The routing table is broadcasted to neighbours after entries have been changed or added as a result of statement execution or incorporation of routes from neighbour messages. Hence, except when many locations have been updated in a short interval, we expect routing tables to be up-to-date or nearly so, taking into account the network size restrictions of the simulator.

Network topology setup and program loading is handled by scripting on top of a custom simple command-line interface (CLI). When starting up, a node controller is assigned a migration procedure through the CLI, which is the same for all node controllers in the network. One desirable feature that is not implemented is CLI control of link characteristics, such as delays.

By default, the simulator starts the initial task on a single startup node. The initial task is defined by the statements in the mandatory starting block of the ABS program. In all our programs, this task creates all the objects used for the duration of the program. Migration and logging does not commence until a method with the name \texttt{setupFinished} is called on some object. There are several reasons for this kind of initialization: it is easier to predict load balancing behaviour with a fixed set of objects, and it is problematic to create new objects on the fly without garbage collection, which we have not implemented.

\subsection{Scenarios}
\label{sec:scenarios}
A network configuration determines the size and topology of the network; large and dense networks give more overhead in the form of routing and load messaging, making simulations slower. Currently, highly connected topologies with in the order of 25 network nodes can be simulated in reasonable time. On this note, we limit the evaluation to networks with three distinct underlying network topologies from sparsely to fully connected: grids, hypergraphs and full meshes. Our base initial setup for each topology has 32 nodes. Since the simulator scales to at least in the order of 100 nodes for sparsely connected topologies, we also investigate grids larger than 32 nodes to compare results.

For defining object behaviour, we have developed a number of ABS programs specifically to run in our simulator. All programs have a setup phase, where a fixed number of objects are initialized, and a phase where the generated objects perform some computation, possibly involving communication; there are no short-lived dynamically created objects. For all programs but one, which implements the Chord distributed hash table (DHT) algorithm \cite{Stoica01}, communication patterns among generated objects follow straightforwardly from the code. This makes it easier to follow what happens during a simulation and to reason about how far an allocation of objects to nodes is from the optimum, factors which we considered particularly important in scenario development. After running initial simulations, we have adjusted parameters in our programs, and in some cases added functionally redundant instructions to get constant and consistent load and messaging, since our migration procedures consider mainly objects with active tasks. With spurious activity among nodes, messaging and load varies greatly, and progress becomes hard to discern. The programs below are available online \cite{abs-net}:

\begin{description}
\item[\texttt{IndependentTasks.abs}] 
  The starting task generates objects, and each generated object is called upon to perform a long-running task. There is no communication among workers---only briefly at startup between the coordinator object, which initializes and assigns tasks, and the generated objects. Since there is no communication, an optimal allocation is an even distribution of objects among nodes, regardless of the network topology.
\item[\texttt{Star.abs}] 
  An object star configuration consists of one center object and one or more fringe objects. The fringe objects in the star continually communicate with the center object, but not among themselves. The program builds a number of independent object star configurations.
\item[\texttt{Ring.abs}]
  The starting task generates objects which know the identifiers of the next object in the ring. The last object generated gets the identifier of the first object. The first object, when called, calls its next object, and so on, until the object which has the first object as next object is reached. In the computation phase, many such calls traverse the ring simultaneously.
\item[\texttt{ChordDHT.abs}]
  An implementation of the Chord DHT algorithm. Key-value mappings are distributed between a number of objects, which all support a put/get interface to clients. Objects are arranged in a ring, but aside from references to their neighbours, each object has $\log(n)$ ``fingers'', references to non-adjacent objects, where $n$ is the size of the keyspace. The addition, or join, of an object to the ring places the new object at a particular position based on its identifier and can trigger global reconfiguration of the ring. During setup, 128 objects are joined to the chord, and each object becomes associated with either a producer object, which continually puts values into the DHT, or a consumer object, which continually attempts to retrieve values from the DHT using pseudorandom keys.
\end{description}

We consider only migration procedures that as a first priority balance out load evenly among nodes in the network. As a consequence, a simulated node controller continually informs neighbour nodes of its load when appropriate, and receives load messages from neighbours in turn, regardless of the migration procedure used. In the simulator, each migration procedure defines a callback method which takes the affected node controller as a parameter. The callback method is invoked, and can result in the migration of several objects to neighbour nodes. The migration procedures used are described below.

\begin{description}
\item[Berenbrink et al.] An adapted version of the distributed load balancing algorithm by Berenbrink et al. \cite{Berenbrink07}, which does not allow neutral moves. One notable difference in the simulator implementation from the abstract description given in Algorithm~\ref{alg:berenbrink} is that only a fixed small number of objects (20) have the possibility to migrate in each cycle, because of limits on the sizes of message buffers.
\item[Berenbrink et al. with neutral moves]
  An adapted version of the distributed load balancing algorithm by Berenbrink et al., which does allow neutral moves, and therefore converges more slowly. The only difference from Algorithm~\ref{alg:berenbrink} is that the if-condition is $l > l'$ instead of $l > l' + 1$. As determined experimentally, only migrating one or two objects per node per cycle leads to significantly less oscillation of objects, compared to when migrating three or more.
\item[Berenbrink et al. with communication intensity]
  A variant of the preceding procedure, where objects are selected for migration based on their affinity to the (randomly) chosen neighbour node, as determined by their communication history with objects in the neighbour node's direction. The communication history is a list of other objects that a given object has communicated with recently, as given by abstract object-local time, defined by the number of tasks finished since initialization. The affinity of an object to the neighbour node is then quantified as the number of objects in the communication history that are located in the direction of the node, according to the routing table.
\item[Weighted neighbour load difference]
  Once every cycle, an object and an adjacent node are chosen uniformly at random and independently. Then, a probability of migration is calculated and enacted based on the difference in load between the current node and the chosen node, with probability 1 for a difference of 10 or more, and probability 0 for a negative difference. If the load difference is $d$, the migration probability becomes $\frac{d}{10}$, adjusted to closest number in the interval $[0, 1]$.
\item[Weighted neighbour load difference with communication intensity]
  Given a randomly chosen object and adjacent node as in the previous procedure, we define the probability of migration according to communication intensity as the number of entries in the object's communication history found in the direction of the node, divided by the total number of entries in the history. This probability is then combined via weighted averaging with the neighbour load difference probability to define the weighted neighbour load with communication procedure. We have used the weight $0.2$ for the communication intensity probability and $0.8$ for the neighbour's load probability.
\end{description}

\begin{algorithm}
\caption{Berenbrink et al. load balancing cycle.}
\label{alg:berenbrink}
\begin{algorithmic}
\FOR{\textbf{each} active object $o$}
\STATE let $u'$ be a neighbour chosen uniformly at random
\STATE let $l$ be the current load, let $l'$ be the last known load of $u'$
\STATE \textbf{if} $l > l' + 1$ \textbf{then} send $o$ to $u'$ with probability $1 - l'/l$
\ENDFOR
\end{algorithmic}
\end{algorithm}

\subsection{Scenario Objectives}
Since our primary objective is to balance node load evenly, we record the load of all individual nodes over time, and then compute the maximum load and load standard deviation. For scenarios with little to no object communication, these are the only measures that are relevant with respect to our objectives. For scenarios with significant messaging, we also consider the number of object-related messages sent (i.e., $\textsc{Call}$ and $\textsc{Future}$ messages) by each node between sampling intervals---with the average number of messages and standard deviation shown. We do not count messages sent by a node to itself via the self-loop arc, since such messages need not go through a physical link in an implementation.

We sample the required quantities from simulations at a fixed global rate, corresponding roughly to a certain number of transitions (1000) in the semantics with imposed fairness via round-robin scheduling.

\subsection{Results}
In this section, we describe simulation results for the scenarios given above.
\subsubsection{Simulations of \texttt{IndependentTasks.abs}}
The program creates 201 objects in total: one starting object which becomes inactive after initialization and 200 objects that each have a task that runs for the course of the program.

As expected, the algorithm by Berenbrink et al. without neutral moves converges very quickly and stays unchanged with no migrations after reaching a state where neighbour load differences are at most one, which on a full mesh is always balanced. For most of the runs on a 32-node hypergraph network topology, the stable state coincided with a completely balanced allocation, or very closely so. For the grid case, the stable allocation in almost all cases deviated significantly from a fully balanced one.

The algorithm variant with neutral moves and two migrations per cycle converges to an almost-stable state quite quickly on a hypergraph, but continues to have minor oscillation of objects. With the same algorithm and five object migrations allowed per cycle, there is considerably more oscillation going on after coming close to a balanced allocation. On a grid topology, where a stable allocation can be further away from a balanced allocation, allowing neutral moves gives better results than disallowing them, as expected. For a grid, the gain from using neutral moves is most distinct in a lower standard deviation compared to the algorithm without neutral moves.

\subsubsection{Simulations of \texttt{Star.abs}}
In the star program, stars are constructed so that each node can hold a whole star, and there is precisely one star per network node. In an optimal allocation, therefore, there are no node-to-node message exchanges at all; all messages are sent locally. 

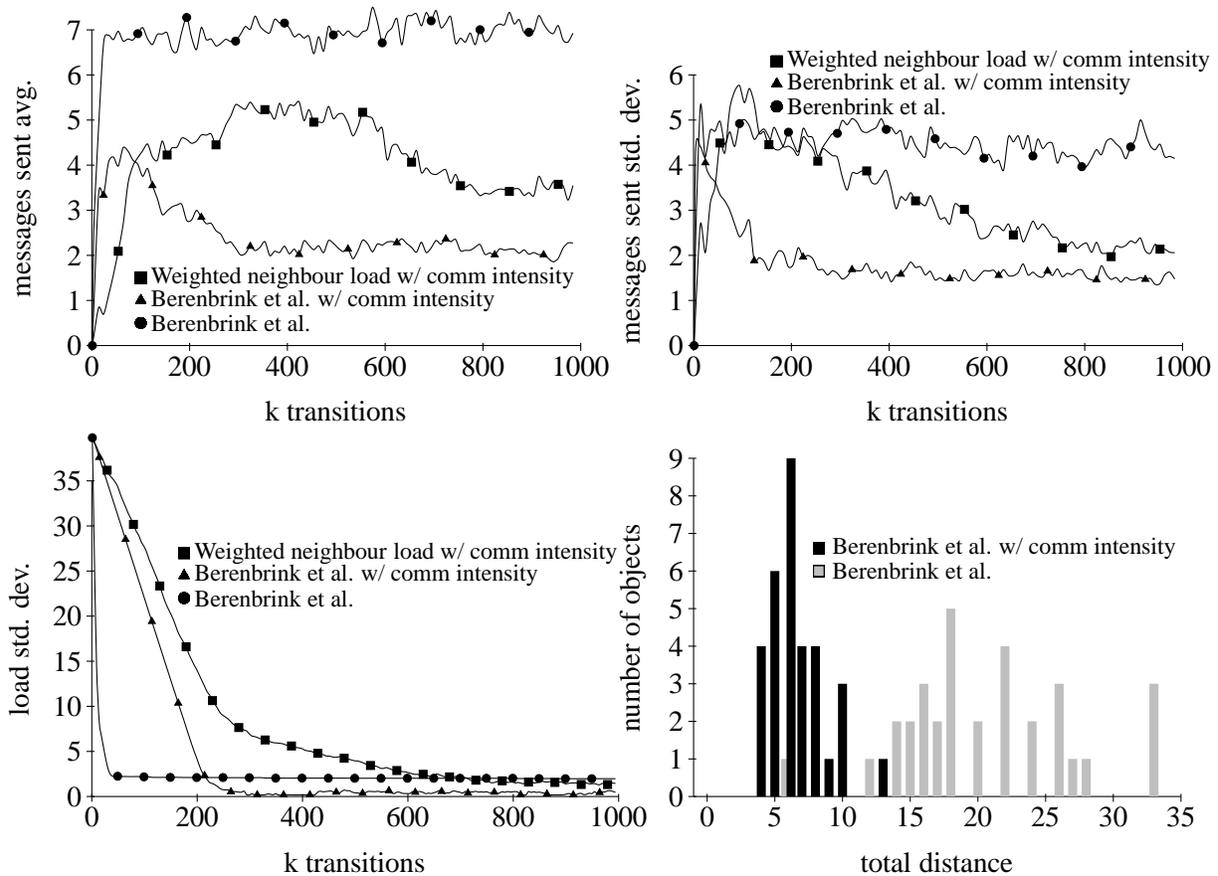
\begin{figure}[ht!]

  \begin{tikzpicture}

    \begin{scope}[xshift=8cm,x=0.0065cm,y=0.6cm]

  \def\xmin{0}
  \def\xmax{1000}
  \def\ymin{0}
  \def\ymax{6}

  % axes
  \draw (\xmin,\ymin) -- coordinate (x axis mid) (\xmax,\ymin);
  \draw (\xmin,\ymin) -- coordinate (y axis mid) (\xmin,\ymax);
 
  % xticks and yticks
  \foreach \x in {0,200,...,1000}
    \draw (\x,1pt) -- (\x,-2pt)
    node at (\x, \ymin) [below] {\x};
  \foreach \y in {0,1,...,6}
    \draw (1pt,\y) -- (-2pt,\y)
    node at (\xmin,\y) [left] {\y};

  % plot the data from the file data.dat
  % smooth the curve and mark the data point with a dot
  \draw plot[smooth] file {data/star_grid_berenbrink_msg_std_smo.dat};
  \draw plot[mark=*, mark size=1.5pt, mark options={fill=black}, only marks] file {data/star_grid_berenbrink_msg_std_smo_marks.dat};  

  \draw plot[smooth] file {data/star_grid_berenbrinkcommunicationintensity_msg_std_smo.dat};
  \draw plot[mark=triangle*, mark size=1.5pt, mark options={fill=black}, only marks] file {data/star_grid_berenbrinkcommunicationintensity_msg_std_smo_marks.dat};

  \draw plot[smooth] file {data/star_grid_communicationintensitytimeout_msg_std_smo.dat};
  \draw plot[mark=square*, mark size=1.5pt, mark options={fill=black}, only marks] file {data/star_grid_communicationintensitytimeout_msg_std_smo_marks.dat};

  %labels      
   \node[below=0.6cm] at (x axis mid) {k transitions};
   \node[rotate=90, above=0.5cm] at (y axis mid) {messages sent std. dev.};

  %legend
   \begin{scope}[shift={(170,5.3)}] 
     \draw (0,0) -- 
     plot[mark=*, mark options={fill=black}] (0.25,0) -- (0.5,0) 
     node[right]{\footnotesize{Berenbrink et al.}};
     \draw[yshift=3mm] (0,0) -- 
     plot[mark=triangle*, mark options={fill=black}] (0.25,0) -- (0.5,0)
     node[right]{\footnotesize{Berenbrink et al. w/ comm intensity}};
     \draw[yshift=6mm] (0,0) -- 
     plot[mark=square*, mark options={fill=black}] (0.25,0) -- (0.5,0)
     node[right]{\footnotesize{Weighted neighbour load w/ comm intensity}};
   \end{scope}

    \end{scope}

    \begin{scope}[x=0.0065cm,y=0.6cm]

  \def\xmin{0}
  \def\xmax{1000}
  \def\ymin{0}
  \def\ymax{7}

  % axes
  \draw (\xmin,\ymin) -- coordinate (x axis mid) (\xmax,\ymin);
  \draw (\xmin,\ymin) -- coordinate (y axis mid) (\xmin,\ymax);

  % xticks and yticks
  \foreach \x in {0,200,...,1000}
    \draw (\x,1pt) -- (\x,-2pt)
    node at (\x, \ymin) [below] {\x};
  \foreach \y in {0,1,...,7}
    \draw (1pt,\y) -- (-2pt,\y)
    node at (\xmin,\y) [left] {\y};

  % plot the data from the file data.dat
  % smooth the curve and mark the data point with a dot
  \draw plot[smooth] file {data/star_grid_berenbrink_msg_avg_smo.dat};
  \draw plot[mark=*, mark size=1.5pt, mark options={fill=black}, only marks] file {data/star_grid_berenbrink_msg_avg_smo_marks.dat};  

  \draw plot[smooth] file {data/star_grid_berenbrinkcommunicationintensity_msg_avg_smo.dat};
  \draw plot[mark=triangle*, mark size=1.5pt, mark options={fill=black}, only marks] file {data/star_grid_berenbrinkcommunicationintensity_msg_avg_smo_marks.dat};

  \draw plot[smooth] file {data/star_grid_communicationintensitytimeout_msg_avg_smo.dat};
  \draw plot[mark=square*, mark size=1.5pt, mark options={fill=black}, only marks] file {data/star_grid_communicationintensitytimeout_msg_avg_smo_marks.dat};

  %labels      
   \node[below=0.6cm] at (x axis mid) {k transitions};
   \node[rotate=90, above=0.6cm] at (y axis mid) {messages sent avg.};

  %legend
   \begin{scope}[shift={(100,0.5)}] 
     \draw (0,0) -- 
     plot[mark=*, mark options={fill=black}] (0.25,0) -- (0.5,0) 
     node[right]{\footnotesize{Berenbrink et al.}};
     \draw[yshift=3mm] (0,0) -- 
     plot[mark=triangle*, mark options={fill=black}] (0.25,0) -- (0.5,0)
     node[right]{\footnotesize{Berenbrink et al. w/ comm intensity}};
     \draw[yshift=6mm] (0,0) -- 
     plot[mark=square*, mark options={fill=black}] (0.25,0) -- (0.5,0)
     node[right]{\footnotesize{Weighted neighbour load w/ comm intensity}};
   \end{scope}

    \end{scope}

    \begin{scope}[yshift=-6cm,x=0.007cm,y=0.12cm]

   \def\xmin{0}
  \def\xmax{1000}
  \def\ymin{0}
  \def\ymax{35}

  % axes
  \draw (\xmin,\ymin) -- coordinate (x axis mid) (\xmax,\ymin);
  \draw (\xmin,\ymin) -- coordinate (y axis mid) (\xmin,\ymax);
 
  % xticks and yticks
  \foreach \x in {0,200,...,1000}
    \draw (\x,1pt) -- (\x,-2pt)
    node at (\x, \ymin) [below] {\x};
  \foreach \y in {0,5,...,35}
    \draw (1pt,\y) -- (-2pt,\y)
    node at (\xmin,\y) [left] {\y};

  % plot the data from the file data.dat
  % smooth the curve and mark the data point with a dot
  \draw plot[smooth] file {data/star_grid_berenbrink_std_smo.dat};
  \draw plot[mark=*, mark size=1.5pt, mark options={fill=black}, only marks] file {data/star_grid_berenbrink_std_smo_marks.dat};  

  \draw plot[smooth] file {data/star_grid_berenbrinkcommunicationintensity_std_smo.dat};
  \draw plot[mark=triangle*, mark size=1.5pt, mark options={fill=black}, only marks] file {data/star_grid_berenbrinkcommunicationintensity_std_smo_marks.dat};

  \draw plot[smooth] file {data/star_grid_communicationintensitytimeout_std_smo.dat};
  \draw plot[mark=square*, mark size=1.5pt, mark options={fill=black}, only marks] file {data/star_grid_communicationintensitytimeout_std_smo_marks.dat};

   %labels
   \node[below=0.6cm] at (x axis mid) {k transitions};
   \node[rotate=90, above=0.7cm] at (y axis mid) {load std. dev.};

  %legend
   \begin{scope}[shift={(175,22)}] 
     \draw (0,0) -- 
     plot[mark=*, mark options={fill=black}] (0.25,0) -- (0.5,0) 
     node[right]{\footnotesize{Berenbrink et al.}};
     \draw[yshift=3mm] (0,0) -- 
     plot[mark=triangle*, mark options={fill=black}] (0.25,0) -- (0.5,0)
     node[right]{\footnotesize{Berenbrink et al. w/ comm intensity}};
     \draw[yshift=6mm] (0,0) -- 
     plot[mark=square*, mark options={fill=black}] (0.25,0) -- (0.5,0)
     node[right]{\footnotesize{Weighted neighbour load w/ comm intensity}};
   \end{scope}
      
    \end{scope}
    
  \begin{scope}[xshift=8cm,yshift=-6cm,x=0.18cm,y=0.5cm]
  \def\xmin{0}
  \def\xmax{36}
  \def\ymin{0}
  \def\ymax{9}

  % axes
  \draw (\xmin,\ymin) -- coordinate (x axis mid) (\xmax,\ymin);
  \draw (\xmin,\ymin) -- coordinate (y axis mid) (\xmin,\ymax);

  % xticks and yticks
  \draw (1,1pt) -- (1,-2pt) node at (1, \ymin) [below] {0};
  \draw (6,1pt) -- (6,-2pt) node at (6, \ymin) [below] {5};
  \draw (11,1pt) -- (11,-2pt) node at (11, \ymin) [below] {10};
  \draw (16,1pt) -- (16,-2pt) node at (16, \ymin) [below] {15};
  \draw (21,1pt) -- (21,-2pt) node at (21, \ymin) [below] {20};
  \draw (26,1pt) -- (26,-2pt) node at (26, \ymin) [below] {25};
  \draw (31,1pt) -- (31,-2pt) node at (31, \ymin) [below] {30};
  \draw (36,1pt) -- (36,-2pt) node at (36, \ymin) [below] {35};

  \foreach \y in {0,...,9} \draw (1pt,\y) -- (-2pt,\y) node at (\xmin,\y) [left] {\y};
 
  \draw[ycomb, color=lightgray,line width=0.12cm] plot file {data/star_grid_berenbrink_dists.dat};
  \draw[ycomb, color=black,line width=0.12cm] plot file {data/star_grid_berenbrinkcommunicationintensity_dists.dat};

  %labels      
  \node[below=0.6cm] at (x axis mid) {total distance};
  \node[rotate=90, above=0.5cm] at (y axis mid) {number of objects};
  
  %legend
  \begin{scope}[shift={(9, 6)}]
    \draw (0,0) -- plot[mark=square*, mark options={fill=lightgray}] (0.25,0) -- (0.5,0) 
    node[right]{\footnotesize{Berenbrink et al.}};
    \draw[yshift=3mm] (0,0) -- plot[mark=square*, mark options={fill=black}] (0.25,0) -- (0.5,0)
    node[right]{\footnotesize{Berenbrink et al. w/ comm intensity}};
  \end{scope}

  \end{scope}

\end{tikzpicture}

\caption{\texttt{Star.abs} running on a 32-node grid.}
\label{fig:star}
\end{figure}

We expected the pure load balancing procedures to have markedly worse results than the procedures taking inter-object communication intensity into account. The average number of sent messages and the standard deviation of sent messages over time for the star program on a grid is shown in the upper half of Figure~\ref{fig:star}, with measurements smoothed out via averaging over five samples to reduce noise. As can be seen, there is a distinct improvement with respect to messages sent when using the algorithm by Berenbrink et al. augmented with message intensity comparisons when compared to the other procedures, although it is quite far from the optimum. The algorithm using probabilistic weighting of load and messaging seems to improve the most over time, although it performs similarly to the messaging-augmented load balancing algorithm by Berenbrink et al.

With all the tested migration strategies for a grid, load became evenly balanced relatively quickly, as seen in the lower left part of Figure~\ref{fig:star}. Hence, there was no significant avoidance of messaging by communicating objects clustering at a few specific nodes.

Because of the simplicity of the object communication graph and the fact that it is possible to reach an allocation where no inter-node communication takes place, it is worthwhile to illustrate how near specific algorithms can get after many (1000) cycles, for comparison. In a given allocation, each object has a total distance in hops to the other object it communicates with. For fringe objects, the total distance is the number of hops to its center object, but center objects have total distance equal to the sum of all distances to its fringes. In an optimal allocation, all centers (and all fringes) have total distance zero. In the lower right part of Figure~\ref{fig:star}, gray bars show the distribution of total distance among the 32 center objects on a grid for the load balancing algorithm by Berenbrink et al. The black bars show the distribution of total distances of the objects for the algorithm by Berenbrink et al. augmented with message intensity comparisons. The distributions intersect, but the former algorithm fares worse.

Results for \texttt{Star.abs} on a hypergraph topology give a less pronounced advantage to the two migration procedures which take message intensity into account. Of those procedures, the Berenbrink et al. variant produces the least messaging, but trends are largely the same as for the grid case; hence, we omit plots. For the case of a complete topology, the amount of messaging was virtually the same for all procedures. An intuition for why this is the case is that it becomes much harder to improve upon an allocation in a situation where migrations are helpful only when communicating objects end up on the same node, and there is additionally no corresponding loss of proximity to another object. Grids of 64 and 128 nodes have the same messaging trends as the 32-node case.

\subsubsection{Simulations of \texttt{Ring.abs}}
When running a ring of 128 objects on a 32-node grid, there are balanced allocations with all nodes having 4 objects, where all objects that communicate are either on the same node or adjacent nodes. The idea is that two of the objects on a node are part of a segment of the ring, while the other two are part of another segment coming back the other way. In such allocations, at most one inter-node message per object is needed for a method invocation that involves the whole ring.

The upper left half of Figure~\ref{fig:ring} shows the average number of messages sent of a 128-object ring on a grid topology, while the upper right half shows the standard deviation of the number of sent messages; smoothing by averaging samples has been applied in both cases. The pattern from the star program remains, with procedures taking messaging into account performing better, but the differences are smaller. The progressively decreasing number of inter-node messages sent are not due to clustering of many objects on a few nodes, as shown by the eventually low standard deviation of load in the lower left part of the figure.

In the lower right part of Figure~\ref{fig:ring}, gray bars show the distribution of total distance among all ring objects on a grid to the objects they communicate with, after 1000 migration cycles using the algorithm by Berenbrink et al. Black bars show the distribution for the algorithm by Berenbrink et al. with neutral moves augmented with message intensity comparisons. There is overlap, but the latter algorithm results in many more objects with total distance between 1 and 5. However, both distributions are quite far from being optimal.

\begin{figure}[ht!]

\begin{tikzpicture}

\begin{scope}[xshift=8cm,x=0.006cm,y=0.8cm]

  \def\xmin{0}
  \def\xmax{1000}
  \def\ymin{0}
  \def\ymax{4}

  % axes
  \draw (\xmin,\ymin) -- coordinate (x axis mid) (\xmax,\ymin);
  \draw (\xmin,\ymin) -- coordinate (y axis mid) (\xmin,\ymax);
 
  % xticks and yticks
  \foreach \x in {0,200,...,1000}
    \draw (\x,1pt) -- (\x,-2pt)
    node at (\x, \ymin) [below] {\x};
  \foreach \y in {0,1,...,4}
    \draw (1pt,\y) -- (-2pt,\y)
    node at (\xmin,\y) [left] {\y};

  % plot the data from the file data.dat
  % smooth the curve and mark the data point with a dot
  \draw plot[smooth] file {data/ring_grid_berenbrink_msg_std_smo.dat};
  \draw plot[mark=*, mark size=1.5pt, mark options={fill=black}, only marks] file {data/ring_grid_berenbrink_msg_std_smo_marks.dat};

  \draw plot[smooth] file {data/ring_grid_berenbrinkcommunicationintensity_msg_std_smo.dat};
  \draw plot[mark=triangle*, mark size=1.5pt, mark options={fill=black}, only marks] file {data/ring_grid_berenbrinkcommunicationintensity_msg_std_smo_marks.dat};

  \draw plot[smooth] file {data/ring_grid_communicationintensitytimeout_msg_std_smo.dat};
  \draw plot[mark=square*, mark size=1.5pt, mark options={fill=black}, only marks] file {data/ring_grid_communicationintensitytimeout_msg_std_smo_marks.dat};

  %labels      
   \node[below=0.6cm] at (x axis mid) {k transitions};
   \node[rotate=90, above=0.5cm] at (y axis mid) {messages sent std. dev.};

  %legend
   \begin{scope}[shift={(180,2.5)}] 
     \draw (0,0) -- 
     plot[mark=*, mark options={fill=black}] (0.25,0) -- (0.5,0) 
     node[right]{\footnotesize{Berenbrink et al.}};
     \draw[yshift=3mm] (0,0) -- 
     plot[mark=triangle*, mark options={fill=black}] (0.25,0) -- (0.5,0)
     node[right]{\footnotesize{Berenbrink et al. w/ comm intensity}};
     \draw[yshift=6mm] (0,0) -- 
     plot[mark=square*, mark options={fill=black}] (0.25,0) -- (0.5,0)
     node[right]{\footnotesize{Weighted neighbour load w/ comm intensity}};
   \end{scope}

  \end{scope}

  \begin{scope}[x=0.006cm,y=0.8cm]

  \def\xmin{0}
  \def\xmax{1000}
  \def\ymin{0}
  \def\ymax{4}

  % axes
  \draw (\xmin,\ymin) -- coordinate (x axis mid) (\xmax,\ymin);
  \draw (\xmin,\ymin) -- coordinate (y axis mid) (\xmin,\ymax);

  % xticks and yticks
  \foreach \x in {0,200,...,1000}
    \draw (\x,1pt) -- (\x,-2pt)
    node at (\x, \ymin) [below] {\x};
  \foreach \y in {0,1,...,4}
    \draw (1pt,\y) -- (-2pt,\y)
    node at (\xmin,\y) [left] {\y};

  % plot the data from the file data.dat
  % smooth the curve and mark the data point with a dot
  \draw plot[smooth] file {data/ring_grid_berenbrink_msg_avg_smo.dat};
  \draw plot[mark=*, mark size=1.5pt, mark options={fill=black}, only marks] file {data/ring_grid_berenbrink_msg_avg_smo_marks.dat};  

  \draw plot[smooth] file {data/ring_grid_berenbrinkcommunicationintensity_msg_avg_smo.dat};
  \draw plot[mark=triangle*, mark size=1.5pt, mark options={fill=black}, only marks] file {data/ring_grid_berenbrinkcommunicationintensity_msg_avg_smo_marks.dat};

  \draw plot[smooth] file {data/ring_grid_communicationintensitytimeout_msg_avg_smo.dat};
  \draw plot[mark=square*, mark size=1.5pt, mark options={fill=black}, only marks] file {data/ring_grid_communicationintensitytimeout_msg_avg_smo_marks.dat};

  %labels
   \node[below=0.6cm] at (x axis mid) {k transitions};
   \node[rotate=90, above=0.5cm] at (y axis mid) {messages sent avg.};

  %legend
   \begin{scope}[shift={(170,2.5)}] 
     \draw (0,0) -- 
     plot[mark=*, mark options={fill=black}] (0.25,0) -- (0.5,0) 
     node[right]{\footnotesize{Berenbrink et al.}};
     \draw[yshift=3mm] (0,0) -- 
     plot[mark=triangle*, mark options={fill=black}] (0.25,0) -- (0.5,0)
     node[right]{\footnotesize{Berenbrink et al. w/ comm intensity}};
     \draw[yshift=6mm] (0,0) -- 
     plot[mark=square*, mark options={fill=black}] (0.25,0) -- (0.5,0)
     node[right]{\footnotesize{Weighted neighbour load w/ comm intensity}};
   \end{scope}     

   \end{scope}

   \begin{scope}[yshift=-5cm,x=0.006cm,y=0.15cm]
   
  \def\xmin{0}
  \def\xmax{1000}
  \def\ymin{0}
  \def\ymax{25}

  % axes
  \draw (\xmin,\ymin) -- coordinate (x axis mid) (\xmax,\ymin);
  \draw (\xmin,\ymin) -- coordinate (y axis mid) (\xmin,\ymax);
 
  % xticks and yticks
  \foreach \x in {0,200,...,1000}
    \draw (\x,1pt) -- (\x,-2pt)
    node at (\x, \ymin) [below] {\x};
  \foreach \y in {0,5,...,25}
    \draw (1pt,\y) -- (-2pt,\y)
    node at (\xmin,\y) [left] {\y};

  % plot the data from the file data.dat
  % smooth the curve and mark the data point with a dot
  \draw plot[smooth] file {data/ring_grid_berenbrink_std_smo.dat};
  \draw plot[mark=*, mark size=1.5pt, mark options={fill=black}, only marks] file {data/ring_grid_berenbrink_std_smo_marks.dat};  

  \draw plot[smooth] file {data/ring_grid_berenbrinkcommunicationintensity_std_smo.dat};
  \draw plot[mark=triangle*, mark size=1.5pt, mark options={fill=black}, only marks] file {data/ring_grid_berenbrinkcommunicationintensity_std_smo_marks.dat};

  \draw plot[smooth] file {data/ring_grid_communicationintensitytimeout_std_smo.dat};
  \draw plot[mark=square*, mark size=1.5pt, mark options={fill=black}, only marks] file {data/ring_grid_communicationintensitytimeout_std_smo_marks.dat};

   %labels
   \node[below=0.6cm] at (x axis mid) {k transitions};
   \node[rotate=90, above=0.6cm] at (y axis mid) {load std. dev.};

  %legend
   \begin{scope}[shift={(160,12)}] 
     \draw (0,0) -- 
     plot[mark=*, mark options={fill=black}] (0.25,0) -- (0.5,0) 
     node[right]{\footnotesize{Berenbrink et al.}};
     \draw[yshift=3mm] (0,0) -- 
     plot[mark=triangle*, mark options={fill=black}] (0.25,0) -- (0.5,0)
     node[right]{\footnotesize{Berenbrink et al. w/ comm intensity}};
     \draw[yshift=6mm] (0,0) -- 
     plot[mark=square*, mark options={fill=black}] (0.25,0) -- (0.5,0)
     node[right]{\footnotesize{Weighted neighbour load w/ comm intensity}};
   \end{scope}

   \end{scope}

   \begin{scope}[xshift=8cm,yshift=-5cm,x=0.35cm,y=0.14cm]

  \def\xmin{0}
  \def\xmax{16}
  \def\ymin{0}
  \def\ymax{30}

  % axes
  \draw (\xmin,\ymin) -- coordinate (x axis mid) (\xmax,\ymin);
  \draw (\xmin,\ymin) -- coordinate (y axis mid) (\xmin,\ymax);

  % xticks and yticks
  \draw (1,1pt) -- (1,-2pt) node at (1, \ymin) [below] {0};
  \draw (6,1pt) -- (6,-2pt) node at (6, \ymin) [below] {5};
  \draw (11,1pt) -- (11,-2pt) node at (11, \ymin) [below] {10};
  \draw (16,1pt) -- (16,-2pt) node at (16, \ymin) [below] {15};
  \foreach \y in {0,5,...,30} \draw (1pt,\y) -- (-2pt,\y) node at (\xmin,\y) [left] {\y};

  \draw[ycomb, color=lightgray,line width=0.15cm] plot file {data/ring_grid_berenbrink_dists.dat};
  \draw[ycomb, color=black,line width=0.15cm] plot file {data/ring_grid_berenbrinkcommunicationintensity_dists.dat};

  %labels      
  \node[below=0.6cm] at (x axis mid) {total distance};
  \node[rotate=90, above=0.6cm] at (y axis mid) {number of objects};
  
  %legend
  \begin{scope}[shift={(5.7,22)}] 
    \draw (0,0) -- plot[mark=square*, mark options={fill=lightgray}] (0.25,0) -- (0.4,0) 
      node[right]{\footnotesize{Berenbrink et al.}};
    \draw[yshift=3mm] (0,0) -- plot[mark=square*, mark options={fill=black}] (0.25,0) -- (0.4,0)
      node[right]{\footnotesize{Berenbrink et al. w/ comm intensity}};
   \end{scope}
     
   \end{scope}

\end{tikzpicture}

\caption{\texttt{Ring.abs} on grid.}
\label{fig:ring}
\end{figure}

As in the case of \texttt{Star.abs}, the performance trend in messaging over time is largely the same on a grid and hypergraph topology for \texttt{Ring.abs}. The main difference on a hypergraph is that procedures which take message intensity into account result in less pronounced improvements over the pure load balancing procedure. For a complete topology, differences are once again small, but with an edge towards the message intensity procedures. Once more, grids of 64 and 128 nodes preserve the trend from the 32-node case.

\subsubsection{Simulations of \texttt{ChordDHT.abs}}
In the Chord DHT program, the weighted neighbour's load and message intensity strategy exhibited a tendency to quickly cause message buffer overflows, while the procedures based on the algorithms of Berenbrink et al. worked largely as expected. 

The left part of Figure~\ref{fig:chorddht} shows the average number of messages sent for nodes when running the program on a grid, and the right part shows the standard deviation of the number of messages. Again, smoothing by averaging samples five at a time has been applied. The results suggest that there is a reasonable payoff from taking messaging into account in a migration strategy, even when running a program with relatively complex communication patterns.

\begin{figure}[ht!]
\begin{tikzpicture}

\begin{scope}[xshift=8cm,x=0.0065cm,y=0.7cm]
  \def\xmin{0}
  \def\xmax{1000}
  \def\ymin{0}
  \def\ymax{7}

  % axes
  \draw (\xmin,\ymin) -- coordinate (x axis mid) (\xmax,\ymin);
  \draw (\xmin,\ymin) -- coordinate (y axis mid) (\xmin,\ymax);
 
  % xticks and yticks
  \foreach \x in {0,200,...,1000}
    \draw (\x,1pt) -- (\x,-2pt)
    node at (\x, \ymin) [below] {\x};
  \foreach \y in {0,1,...,7}
    \draw (1pt,\y) -- (-2pt,\y)
    node at (\xmin,\y) [left] {\y};

  % plot the data from the file data.dat
  % smooth the curve and mark the data point with a dot
  \draw plot[smooth] file {data/chorddht_grid_berenbrink_msg_std_smo.dat};
  \draw plot[mark=*, mark size=1.5pt, mark options={fill=black}, only marks] file {data/chorddht_grid_berenbrink_msg_std_smo_marks.dat};  

  \draw plot[smooth] file {data/chorddht_grid_berenbrinkcommunicationintensity_msg_std_smo.dat};
  \draw plot[mark=triangle*, mark size=1.5pt, mark options={fill=black}, only marks] file {data/chorddht_grid_berenbrinkcommunicationintensity_msg_std_smo_marks.dat};

  %labels      
   \node[below=0.6cm] at (x axis mid) {k transitions};
   \node[rotate=90, above=0.5cm] at (y axis mid) {messages sent std. dev.};

  %legend
   \begin{scope}[shift={(150,3)}] 
     \draw (0,0) -- 
     plot[mark=*, mark options={fill=black}] (0.25,0) -- (0.5,0) 
     node[right]{\footnotesize{Berenbrink et al.}};
     \draw[yshift=3mm] (0,0) -- 
     plot[mark=triangle*, mark options={fill=black}] (0.25,0) -- (0.5,0)
     node[right]{\footnotesize{Berenbrink et al. w/ comm intensity}};
   \end{scope}

\end{scope}

\begin{scope}[x=0.0065cm,y=0.7cm]

  \def\xmin{0}
  \def\xmax{1000}
  \def\ymin{0}
  \def\ymax{6}

  % axes
  \draw (\xmin,\ymin) -- coordinate (x axis mid) (\xmax,\ymin);
  \draw (\xmin,\ymin) -- coordinate (y axis mid) (\xmin,\ymax);

  % xticks and yticks
  \foreach \x in {0,200,...,1000}
    \draw (\x,1pt) -- (\x,-2pt)
    node at (\x, \ymin) [below] {\x};
  \foreach \y in {0,1,...,6}
    \draw (1pt,\y) -- (-2pt,\y)
    node at (\xmin,\y) [left] {\y};

  % plot the data from the file data.dat
  % smooth the curve and mark the data point with a dot
  \draw plot[smooth] file {data/chorddht_grid_berenbrink_msg_avg_smo.dat};
  \draw plot[mark=*, mark size=1.5pt, mark options={fill=black}, only marks] file {data/chorddht_grid_berenbrink_msg_avg_smo_marks.dat};  

  \draw plot[smooth] file {data/chorddht_grid_berenbrinkcommunicationintensity_msg_avg_smo.dat};
  \draw plot[mark=triangle*, mark size=1.5pt, mark options={fill=black}, only marks] file {data/chorddht_grid_berenbrinkcommunicationintensity_msg_avg_smo_marks.dat};

  %labels      
   \node[below=0.6cm] at (x axis mid) {k transitions};
   \node[rotate=90, above=0.5cm] at (y axis mid) {messages sent avg.};

  %legend
   \begin{scope}[shift={(180,3.8)}] 
     \draw (0,0) -- 
     plot[mark=*, mark options={fill=black}] (0.25,0) -- (0.5,0) 
     node[right]{\footnotesize{Berenbrink et al.}};
     \draw[yshift=3mm] (0,0) -- 
     plot[mark=triangle*, mark options={fill=black}] (0.25,0) -- (0.5,0)
     node[right]{\footnotesize{Berenbrink et al. w/ comm intensity}};
   \end{scope}

\end{scope}

\end{tikzpicture}
\caption{\texttt{ChordDHT.abs} on grid.}
\label{fig:chorddht}
\end{figure}

Simulations of \texttt{ChordDHT.abs} on a hypergraph show very similar trends in performance to the grid case, but give a less pronounced advantage to procedures which take message intensity into account, as for previous programs. In a fully connected topology, the procedures result in effectively the same amount of messaging, as before.

\section{Conclusions and Future Work}
\label{conclusions}
The simulation results suggest that it is feasible in a decentralized setting to meet the objective of balanced resource allocation, and also make headway towards the objective of minimizing communication of distributed objects. The results also validate the applicability of the ABS-NET model with location-independent routing to decentralized runtime adaptation. The main concern for relevance to real-world networks is the use in the model of unbounded message queues, and the lack of rate limitation and latency controls in our simulator.

In future work, we plan to continue the theoretical and simulation-based studies to deepen our understanding of multi-dimensional resource management, to improve the performance and accuracy of the simulator, and to investigate adaptation in dynamic networks, initially only with benign churn, i.e., with controlled startup and shutdown of nodes.

\section*{Acknowledgements}
We thank the anonymous reviewers for their comments and suggestions, which were of significant help in improving the paper. We also thank our colleagues in the HATS project for useful discussions and criticism.

\bibliographystyle{eptcs}
\bibliography{bibli}

\end{document}